\documentstyle[aps,epsf,preprint]{revtex}
\tighten
\begin{document}
\draft
\preprint{Version 1.00}

\title{Lamb shift of 3P and 4P states and 
the determination of $\alpha$}

\author{U. D. Jentschura$^{1,}$\footnote{Electronic address: ulj@nist.gov.}, 
G. Soff$^{1}$ 
and P. J. Mohr$^{2,}$\footnote{Electronic address: mohr@nist.gov.}}

\address{
$^1$Institut f\"ur Theoretische Physik, TU Dresden,
Mommsenstra\ss e 13,
01062 Dresden,
Germany}

\address{
$^2$Atomic Physics Division,
National Institute of Standards and Technology (NIST),
Gaithersburg, 
Maryland MD 20899-0001, USA}      

\maketitle

\begin{abstract}
The fine structure interval of P states in hydrogenlike systems
can be determined theoretically with high precision,
because the energy levels of P states are only slightly influenced
by the structure of the nucleus. 
Therefore a measurement of the fine structure may serve 
as an excellent test of QED
in bound systems or alternatively as a 
means of determining the fine structure constant $\alpha$ with
very high precision. In this paper  
an improved analytic calculation 
of higher-order binding corrections to the one-loop self energy of 
3P and 4P states in hydrogen-like systems with low nuclear charge 
number $Z$ is presented.
The method of calculation has been described earlier 
\cite{jp1} and is applied here
to the excited P states. Because of the more complicated nature 
of the wave functions and
the bound state poles corresponding to 
decay of the excited states, the calculations
are more complex. Comparison of the analytic
results to the extrapolated numerical data
for high $Z$ ions \cite{mohrkim} serves as an 
independent test of the analytic evaluation.
New theoretical 
values for the Lamb shift of the P states and 
for the fine structure splittings are given.
\end{abstract}

\narrowtext
\section{INTRODUCTION}
Evaluations of the radiative corrections in higher order
for bound states are an
involved task because of the appearance
of a multitude of terms, and because of the difficulties associated
with bound state formalism. In this paper, we present an improved
calculation of higher order corrections to the one--loop self energy 
of an electron in an excited 3P or 4P state. 

For the contribution $\delta E_{\rm SE}$
of the one--loop radiative correction to the Lamb
shift of a bound electron, we have the following non-analytic expansion
in powers of $Z$ times the fine structure constant $\alpha$ 
\begin{equation}
\delta E_{\rm SE} = \frac{\alpha}{\pi} \, \frac{(Z \alpha)^4 \, m}{n^3} \, F\,,
\label{definitionofF}
\end{equation}
where
\begin{eqnarray}
\label{defF}
F &=& A_{4,1} \, \ln(Z \alpha)^{-2} + A_{4,0} + 
  (Z \alpha) \, A_{5,0} +\nonumber \\
& &  (Z \alpha)^2 \left[A_{6,2} \, \ln^2(Z \alpha)^{-2} +
A_{6,1} \,\ln(Z \alpha)^{-2} + A_{6,0} +
     (Z \alpha) \, G_{{\rm SE},7} \right].
\end{eqnarray} 
The remainder function $G_{{\rm SE},7}$ is of order $1$ and
is comprised of the terms $A_{7,0}$ and higher coefficients. 
Corrections $A_{4,1}$, $A_{5,0}$ and $A_{6,2}$ vanish for P
states. The terms $A_{4,0}$ (see, e.g., \cite{sapirsteinyennie1}) 
and $A_{6,1}$ \cite{sapirsteinyennie1,ericksonyennie12} are known
analytically. The term $A_{4,0}$ contains the Bethe logarithm which
has been evaluated to 12 significant figures \cite{km1,ds1}.   
Results have not been obtained for $A_{6,0}$ coefficients.
In this paper, we present an evaluation of the $A_{6,0}$ coefficients
for the $3{\rm P}_{1/2}$, $3{\rm P}_{3/2}$, $4{\rm P}_{1/2}$ and
$4{\rm P}_{3/2}$ states. The results lead to improved values for the
Lamb shift of the respective states and to a new theoretical value for
the fine structure splitting. We give an explicit formula for the fine
structure of the $2{\rm P}$, $3{\rm P}$ and $4{\rm P}$ states as a 
function of the fine structure constant $\alpha$, which can be used to
obtain a value of $\alpha$ from experimental data.

In this paper, we briefly compare some
of the methods that have been developed for the treatment of the
one--loop problem. We give a brief account and 
illustrate the usefulness of the
$\epsilon$-method \cite{jp1,krp1} for analytic evaluations. 
We then describe the evaluation of the high--energy part
to the self energy, with a focus on details of the integration
procedure. We then proceed to the low--energy part.  
Results of the calculation are given,  
specific contributions are discussed in detail.

\section{Various methods of treatment of the one--loop 
self energy}

Using units in which $\hbar = c = 1$ and $e^2 = 4 \pi \alpha$,
we can write the integral corresponding to the one--loop 
self energy of an electron bound in a Coulomb field, 
\begin{eqnarray}
\label{deltaESE}
\delta E_{\rm SE} &=& \lim_{M\to\infty}
-i e^2 \int_{C_F} \frac{d\omega}{2 \pi} 
\int \frac{d^3 {\bf k}}{(2 \pi)^3} \,
D_{\mu,\nu}^{\rm reg}(k^2,M)\,
\langle \bar{\psi} | \gamma^{\mu} 
  \frac{1}{\not{\! p} - \not{\! k} - m - \gamma^0 V} \gamma^{\nu} 
  | \psi \rangle  \nonumber\\ 
&& - \langle \bar{\psi} | \delta m (M) | \psi \rangle,
\end{eqnarray}
where $D_{\mu,\nu}^{\rm reg}(k^2,M)$ is the Pauli-Villars regularized photon
propagator (in Feynman gauge, we have 
$D_{\mu\nu}^{\rm reg}(k^2,M) = - g_{\mu\nu}(1/k^2-1/(k^2-M^2))$. 
The term $\delta m (M)$ in Eq. (\ref{deltaESE}) is the 
one--loop mass counter term as a
function of $M$, $\delta m (M) =  \alpha\,(3/(4\,\pi))\,m\,
(\ln(M^2/m^2)+1/2$). $\bar{\psi} = \psi^{\dagger} \, \gamma^0$ denotes
the Dirac adjoint. It is straightforward 
to derive Eq. (\ref{deltaESE}) with the 
Feynman rules of QED. By rescaling all variables to 
the electron mass scale
\begin{equation}
\omega \to m \, \omega', \quad {\bf k} \to m \, {\bf k'}, 
\quad p \to m\,p', \quad V \to m\,V', \quad M \to m \, M',
\end{equation}
we have
\begin{eqnarray}
\delta E_{\rm SE} &=& - i e^2 \, m \,\int_{C_F} \frac{d\omega'}{2 \pi} 
\int \frac{d^3 {\bf k'}}{(2 \pi)^3} \,
\left[ \frac{1}{k'^2} - 
\frac{1}{k'^2 - M'^2} \right]\,
\langle \bar{\psi} | \gamma^{\mu} 
  \frac{1}{\not{\! p'} - \not{\! k'} - 1 - \gamma^0 V'} \gamma_{\mu} 
  | \psi \rangle  \nonumber\\ 
&& - \langle \bar{\psi} | \delta m (M') | \psi \rangle. 
\end{eqnarray}
We will use variables rescaled to the electron mass in this paper,
and suppress the prime of the rescaled variables in the sequel.
Note that in our system of units, we have (e. g.) for 
the Bohr radius of the atom
$a_{\rm Bohr} = 1/(Z \alpha)$. By contrast, in atomic units,  
which are used for example in \cite{bookbethe},
we would have the Bohr radius of length unity.
 
The analytic properties of the propagators determine the 
location of the poles in
the integrand in Eq. (\ref{deltaESE}) as indicated in Fig. 
\ref{intcontour1}. The original Feynman prescription calls for
integrating the photon energy along the contour $C_F$.
For the actual evaluation of the Lamb shift, however,
a different contour of integration is used by most authors.
Taking advantage of the analytic properties of the integrand and of
Jordan's lemma, one can change the Feynman contour in the complex plane without
changing the result of the calculation. We compare here the 
contour used in Bethe's original 
derivation of the Lamb shift, the contour used by Mohr 
in \cite{mohr1,mohr2,mohr3,mohr4}, 
and the contour used in Pachucki's $\epsilon$ method, which is  
used in this paper. 

\begin{figure}[htb]
\centerline{\mbox{\epsfysize=5.0cm\epsffile{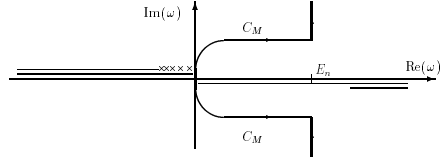}}}
\caption{\label{intcontour2} Mohr's contour for evaluating the
one--loop self energy contribution to the Lamb shift. }
\end{figure}

Mohr's and Pachucki's methods both depend 
on a division of the calculation into a low and a high--energy part.
Mohr's method relies on the contour $C_M$ in Fig.
\ref{intcontour2}. His low--energy part 
is determined by the part of the contour $C_M$ where 
${\rm Re}(\omega) < \epsilon$. 
The residues of the poles of the photon propagator only contribute to 
low--energy part in this case. It can be shown that the low--energy part is given
by the formula
\begin{eqnarray}
\label{mohrlep}
\Delta E_{L} &=& \lim_{\delta \to 0+} \Bigg[
\, \frac{\alpha}{\pi} \, E_n - \frac{\alpha}{4\,\pi^2} 
\, \int_{k < E_n} d^3 k \, \frac{1}{k} 
\left(\delta^{i,j} - \frac{k^i \, k^j}{k^2}\right) \nonumber\\
&& \times \langle{\psi} | \alpha^i \,
e^{i {\bf k}\cdot{\bf r}} \, \frac{1}{H_D - E_n - k - i \delta} \, \alpha^j \,
e^{-i {\bf k}\cdot{\bf r}}| \psi \rangle \Bigg].
\end{eqnarray}
(cf. Eq. (3.8) in \cite{mohr1}, $H_D$ is the Dirac Hamiltonian). 
This contribution contains terms of lower order in 
$(Z \alpha)$ than $(Z \alpha)^4$. The spurious lower order terms cancel when 
the low and the high--energy parts are added in that method. The 
high--energy part is obtained by Wick-rotating the 
Feynman contour for $\omega$ integration 
along the line with ${\rm Re}(\omega) = E_n$. In the non-relativistic limit, 
expression (\ref{mohrlep}) corresponds 
(up to the term $\alpha/\pi \, E_n$) 
to what would be expected to be the self energy of the electron in terms of 
traditional second order perturbation theory due to transverse 
modes of the electromagnetic field, 
\begin{eqnarray}
\label{deltaEtrad}
\Delta E_L^{(2)} &=& {\rm Re}(\Delta E_{L}) - 
\frac{\alpha}{\pi} \, E_n \nonumber\\
&=& \sum_{n'}\!\!\!\!\!\!\!\!\int 
\,\,\, P\,\int_{k < K} d^3 k \sum_{\lambda = 1,2}
\frac{e^2}{4\,m^2} \int d^3 r \frac{|\psi^\dagger_{n'}({\bf x}) \left[
\bbox{\nabla}
\cdot
\bbox{\epsilon}_{\lambda}({\bf k}) \, e^{i {\bf k}\cdot{\bf r}} +
e^{i {\bf k}\cdot{\bf r}} \, \bbox{\nabla} \cdot
\bbox{\epsilon}_{\lambda}({\bf k})\right] \psi_{n'}({\bf x})|^2}
{E_n - E_{n'} - k}, 
\end{eqnarray}
where 
\begin{equation}
\sum_{\lambda=1,2} \epsilon_{\lambda}^{i}({\bf k}) \, 
\epsilon_{\lambda}^{j}({\bf k}) = \delta^{ij} - \frac{k^i\, k^j}{k^2}.
\end{equation}
$K$ in Eq. (\ref{deltaEtrad}) is an appropriate energy cutoff 
to make the expression finite (for a derivation of  Eq. (\ref{deltaEtrad}) cf.
\cite{itzykson}, Eq. (7-112) to Eq. (7-115) ibid., where in the
relativistic case one has to substitute $\bbox{\alpha}$ for 
$1/i \bbox{\nabla}$).
In Mohr's method, $K$ corresponds to $E_n$. 
Bethe's derivation of the Lamb shift, which gave the right scaling of 
the effect of the self energy and correctly identified $A_{4,1}$,
but did not include the contribution to the Lamb shift of order
$\alpha/\pi (Z \alpha)^4$ due to the anomalous magnetic moment, 
comprised the expression given in
Eq. (\ref{deltaEtrad}), but with a major modification. Bethe subtracted
from (\ref{deltaEtrad}) the contribution that would
modify the energy (or mass) of a free electron due 
to its self interaction.
This contribution would give a contribution to 
the rest mass of any electron, and thus would be unobservable. In our
terminology, Bethe's non-relativistic (NR) expression would be written
\begin{eqnarray}
\label{bethe}
\delta E_L^{2, {\rm NR}} = - {\rm P} \,
\frac{2 \alpha}{3 \, \pi} \int_0^{K = m} dk \, k 
\langle \phi | \frac{p^i}{m} \, 
\left[\frac{1}{H_S - (E_\phi - k)} - \frac{1}{k} \right]
\frac{p^i}{m} | \phi \rangle,
\end{eqnarray}
where $H_S$ is the Schr\"odinger Hamiltonian, $\phi$ is the
non-relativistic wave function, and
the subtracted term $- 1/k$ in the integrand corresponds
to the portion of mass renormalization attributable to the low--energy
part. By taking the principal value (P), we identify the real part of
Eq. (\ref{bethe}) as the energy shift, whereas the imaginary part 
corresponds to the decay width of the state $| \phi \rangle$.
Using the subtraction, Bethe disposed of the spurious 
lower-order terms 
and obtained a finite expression.

\begin{figure}[htb]
\centerline{\mbox{\epsfysize=5.0cm\epsffile{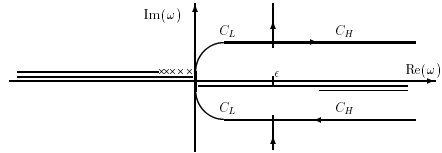}}}
\caption{\label{intcontour3} The $\omega$-integration contour used
by Pachucki and in the calculation presented in this paper. For the 
divergent terms in the high--energy part, we use the Wick-rotated
contour given by the lines extending to $\epsilon  \pm i\,\infty$. For
the naively convergent terms, we use the original contour $C_H$
which extends to $+\infty \pm i \, \delta$. }
\end{figure}

In Pachucki's method (see Fig. \ref{intcontour3}), 
an expression similar to Eq. 
(\ref{deltaEtrad}) is obtained in the 
non-relativistic limit for the low--energy part, but with an
upper cutoff epsilon for the photon energy. This cutoff epsilon separates
the low and the high--energy parts. In the dipole approximation
$\exp(i {\bf k}\cdot {\bf r}) \to 1$, one obtains the expression
\begin{eqnarray}
\delta E_L = - \frac{2 \alpha}{3 \, \pi} \int_0^{\epsilon} dk \, k 
\langle \phi | \frac{p^i}{m} \, \frac{1}{H_S - (E_\phi - k)}
\frac{p^i}{m} | \phi \rangle
\end{eqnarray}
for the low-energy part in leading order.
The renormalization term $- 1/k$ is gone, and the upper
cutoff has been changed from $K=m$ to $K=\epsilon$.
The justification for leaving out the renormalization term
is intimately linked to the special series expansion prescription
used by Pachucki.

Pachucki's method relies on the fact that the 
low--energy part and the high--energy part may formally be regarded 
as functions of the fine structure constant $\alpha$ and the
cutoff parameter $\epsilon$. Their sum, however, the self energy
of the electron $\delta E$,
\begin{equation}
\delta E(\alpha) = E_L(\alpha, \epsilon) + E_H(\alpha, \epsilon),
\end{equation}
does not depend on epsilon, provided the high and the low--energy
parts are expanded first in $\alpha$, and then in epsilon (the order
of expansion plays a crucial role in that case).

Another important point in Pachucki's method is that the spurious lower order
terms which were present in Mohr's calculation vanish in the
limit $\epsilon \to 0$, so we do not need to take them into
account. For example, in Mohr's calculation, the first spurious
term $(\alpha/\pi) \, E_n$ originated from a trivial integration
$\int_0^{E_n} dk \, \alpha/\pi = (\alpha/\pi) \, E_n$. In Pachucki's
method, we would change the upper limit of integration to 
$\epsilon$ and calculate
$\lim_{\epsilon \to 0} \int_0^\epsilon dk \, \alpha/\pi = 0$.
That means by choosing the $\epsilon$ prescription, we not only make the
expression for the low--energy part separately finite, but also dispose
of the spurious lower order terms. That is the principal reason why
Pachucki's method is well suited for the analytic calculation of 
higher order corrections to the one--loop self energy. 

The choice of epsilon
remains arbitrary to a certain extent (it has to be because we 
analytically expand in $\epsilon$ and thus require arbitrariness). However, 
we must put some restraints on the magnitude of $\epsilon$. 
In the high--energy part, we expand the propagator of the bound electron
in powers of the binding field $V$.  
We initially assume a fixed value for $\epsilon$
which prevents infrared problems, but since 
eventually $\epsilon \rightarrow 0$,
this expansion is regarded as a formal expansion that is not necessarily 
convergent.
However, $\epsilon$ may not be 
arbitrarily large. If we let $\epsilon > 2 m$, we enclose poles not only from 
the photon propagator, but also from the negative spectrum of the 
Dirac-Coulomb propagator, which would significantly alter our expression for
the low--energy part. It is also required that in the entire domain of the
low--energy part, an expansion of the expression
\begin{displaymath}
\exp(i {\bf k}\cdot{\bf r})
\end{displaymath}
in the matrix element
\begin{displaymath}
\langle{\psi} | \alpha^i \,
e^{i {\bf k}\cdot{\bf r}} \, \frac{1}{H_D - E_n - k - i \delta} \, \alpha^j \,
e^{-i {\bf k}\cdot{\bf r}}| \psi \rangle
\end{displaymath}
in powers of ${\bf k} \cdot {\bf r}$ corresponds to an 
expansion in powers of $Z \alpha$ (this requirement
justifies the so-called dipole approximation, in which we replace
$\exp(i {\bf k}\cdot{\bf r})$ by unity to obtain the 
lowest order contribution to the self energy).
The order of magnitude of $r$ is $1/(Z \alpha)$ in natural units.
Thus we require $\epsilon < (Z \alpha)$. The dominant contribution 
is then determined by the region in which the photon energy 
$k \equiv \omega = O((Z \alpha)^2)$, so that 
${\bf k}\cdot{\bf r} = O(Z \alpha)$. In this paper we consider
the relativistic corrections up to relative order $(Z \alpha)^2$. This
corresponds to expanding $\exp(i {\bf k}\cdot{\bf r})$ up to 
$({\bf k}\cdot{\bf r})^2$.
Our restrictions on the magnitude of 
$\epsilon$ do not compromise the
validity of analytic expansion in the parameter $\epsilon$.

\section{The high-energy part}

\begin{figure}[htb]
\centerline{\mbox{\epsfysize=5.0cm\epsffile{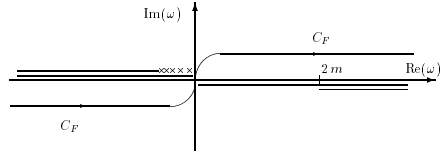}}}
\caption{\label{intcontour1} Feynman contour for $\omega$ integration
(one--loop self energy). Lines
directly below and above the real axis denote
branch cuts from the photon and electron propagator. Crosses denote 
poles originating from the discrete spectrum of the electron 
propagator.}
\end{figure}

\begin{figure}[htb]
\centerline{\mbox{\epsfysize=3.5cm\epsffile{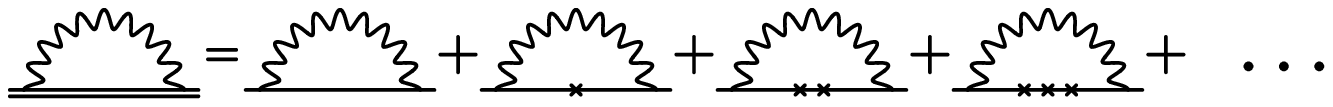}}}
\caption{\label{Vexp} Expansion of the bound electron self energy in
powers of the binding field. }
\end{figure}

The high--energy part of the radiative correction is given by
\begin{eqnarray}
\label{defEH}
E_H &=& - i e^2 m \, \int_{C_H} \frac{d\omega}{2 \pi} 
\int \frac{d^3 {\bf k}}{(2 \pi)^3} \,
\left[ \frac{1}{k^2} - 
\frac{1}{k^2 - M^2} \right]\,
\langle \bar{\psi} | \gamma^{\mu} 
  \frac{1}{\not{\! p} - \not{\! k} - 1 - \gamma^0 V} \gamma_{\mu} 
  | \psi \rangle  \nonumber\\ 
&& - \langle \bar{\psi} | \delta m(M) | \psi \rangle, 
\end{eqnarray}
where we have used the Feynman gauge for the photon propagator 
($D_{\mu \nu}(k) = - g_{\mu \nu} / k^2$)
and the Pauli-Villars regularization prescription
\begin{equation}
\label{paulivillars}
\frac{1}{k^2 + i \delta} \to \frac{1}{k^2 + i \delta} - 
\frac{1}{k^2 - M^2 + i \delta},
\end{equation}
Note that we may leave out $i \epsilon$ prescription
when integrating along $C_H$ since we take the difference of the integrand
infinitesimally above and below the real axis on $C_H$. 
Along the positive real axis,
the integrand has branch cuts due to the photon and electron propagators
as depicted in Fig. \ref{intcontour1}. 
The expression given in Eq. (\ref{defEH}) for $E_H$ is infrared divergent. 
In the evaluation, we start by calculating the matrix element
\begin{equation}
{\tilde P} = \langle \bar{\psi} | \gamma^{\mu} 
  \frac{1}{\not{\! p} - \not{\! k} - m - \gamma^0 V} \gamma_{\mu} 
  | \psi \rangle 
\end{equation}
up to the order of $(Z \alpha)^6$. As outlined in \cite{jp1},
this can be achieved by first expanding the electron propagator
in powers of the binding Coulomb field $V$. This leads to a 3-vertex,
a double-vertex, a single-vertex and a zero-vertex part. The expansion can
be diagrammatically represented as in Fig. \ref{Vexp}. 
The resulting expressions are subsequently expanded
in powers of the spatial electron 
momenta $p^i$. This procedure is feasible for P states because up to order
$(Z \alpha)^6$, all of the resulting matrix elements converge. After
performing the algebra of the Dirac matrices, the resulting 
matrix elements on the P state are evaluated by symbolic 
procedures written in
the computer algebra system {\sc Mathematica} \cite{mathematica}. 
For the evaluation,
we first expand the wave function (given by the exact solution to
the Dirac-Coulomb equation) in powers of $(Z \alpha)$, then we apply
operators in coordinate space representation and finally integrate
the resulting expressions with the help of a set of rules that apply to
standard integrals. The integrands which are to be evaluated for the
matrix elements have lengths of up to 2,000 terms. 
  
We use a parametric representation of the mass counter term
to allow for local cancellation of the divergences. It can be shown
that
\begin{equation}
\delta m(M) = -i e^2 m \, \int_{C_H} \frac{d\omega}{2 \pi} 
\int \frac{d^3 {\bf k}}{(2 \pi)^3} \,
\,\left[\frac{1}{\omega^2- {\bf k}^2}-
\frac{1}{\omega^2- {\bf k}^2 - M^2}\right]\,
\frac{2(\omega + 1)}{\omega^2 - {\bf k}^2 - 2 \,\omega}
\end{equation}
is a suitable parametric representation of the mass counter term
along the contour $C_H$. The portion of mass renormalization along
the contour $C_L$ vanishes in the limit $\epsilon \to 0$, and we have
\begin{eqnarray}
\delta m(M)  &=& -i e^2 m \, \int_{C_F} \frac{d\omega}{2 \pi} 
\int \frac{d^3 {\bf k}}{(2 \pi)^3} \,
\,\left[\frac{1}{k^2+i\epsilon}-
\frac{1}{k^2 - M^2 + i \epsilon}\right]\,
\frac{2(\omega + 1)}{\omega^2 - {\bf k}^2 -2 \,\omega} \nonumber\\
&=& \alpha\,\frac{3\,m}{4\,\pi}\,\left[ \ln\left(M^2\right)+
1/2 \right].
\end{eqnarray}
Therefore, by (locally) subtracting the expression
\begin{displaymath}
\delta m_{l} = \frac{2(\omega+1)}{\omega^2- {\bf k}^2 - 2 \,\omega} \, 
\langle \bar{\psi} | \psi \rangle
\end{displaymath}
before the final 
$d\omega \, d^3 {\bf k} \,\, \left[1/(\omega^2 - {\bf k}^2) -
1/(\omega^2 - {\bf k}^2 - M^2)\right]$-integration in Eq. (\ref{defEH}),
we can subtract the divergences associated with mass renormalization.

Note that $\delta m_{l}$ contains the matrix element 
$\langle \bar{\psi} | \psi \rangle$, which is state dependent. Using the
virial theorem for the Dirac-Coulomb equation ($\langle \bbox{\alpha} \,\cdot\,
{\bf p} \rangle = - \langle V \rangle$), we have 
$\langle \bar{\psi} | \psi \rangle 
= E_{\psi}$, where $E_{\psi}$ is the dimensionless Dirac energy 
of the state $\psi$.

We give here the result for the renormalized matrix element
\begin{equation}
{\tilde P}_{\rm ren} = \tilde P - \delta m_l
\end{equation}
up to $(Z \alpha)^6$ in terms of $\bf{k}$ and $\omega$. We have for the $3P_{1/2}$
state,
\begin{eqnarray}
{\tilde P}_{\rm ren}(3P_{1/2}) &=& 
  (Z \alpha) ^2\,\bigg[ -8\,{\bf k}^2 + 4\,{\bf k}^4 + 6\,{\bf k}^2\,\omega  - 
         3\,{\bf k}^4\,\omega  + 12\,\omega ^2 - 10\,{\bf k}^2\,\omega ^2 \nonumber\\&&
- 
         6\,\omega ^3 + 6\,{\bf k}^2\,\omega ^3 + 6\,\omega ^4 - 
         3\,\omega ^5 \bigg] \bigg/ \bigg[ 
     27\,\bigg( {\bf k}^2 + 2\,\omega  - \omega ^2 \bigg) ^3\bigg] \nonumber\\&&
+ 
   (Z \alpha) ^4\,\bigg[ -128\,{\bf k}^4 + 48\,{\bf k}^6 + 24\,{\bf k}^8 + 
         96\,{\bf k}^4\,\omega  + 2\,{\bf k}^6\,\omega  - 9\,{\bf k}^8\,\omega  \nonumber\\&&
+ 
         256\,{\bf k}^2\,\omega ^2 - 156\,{\bf k}^4\,\omega ^2 - 
         90\,{\bf k}^6\,\omega ^2 - 168\,{\bf k}^2\,\omega ^3 + 
         46\,{\bf k}^4\,\omega ^3 \nonumber\\&&
+ 36\,{\bf k}^6\,\omega ^3 - 
         16\,\omega ^4 + 240\,{\bf k}^2\,\omega ^4 + 
         126\,{\bf k}^4\,\omega ^4 + 72\,\omega ^5 \nonumber\\&&
- 
         98\,{\bf k}^2\,\omega ^5 - 54\,{\bf k}^4\,\omega ^5 - 
         132\,\omega ^6 - 78\,{\bf k}^2\,\omega ^6 + 50\,\omega ^7 + 
         36\,{\bf k}^2\,\omega ^7 \nonumber\\&&
+ 18\,\omega ^8 - 9\,\omega ^9
          \bigg] \bigg/ \bigg[ 
     324\,\bigg( {\bf k}^2 + 2\,\omega  - \omega ^2 \bigg) ^5\bigg] \nonumber\\&&
+ 
   (Z \alpha) ^6\,\bigg[ -1044992\,{\bf k}^6 + 516224\,{\bf k}^8 + 
         319032\,{\bf k}^{10} + 32340\,{\bf k}^{12} \nonumber\\&&
+ 587776\,{\bf k}^4\,\omega  + 
         1716736\,{\bf k}^6\,\omega  + 26320\,{\bf k}^8\,\omega  - 
         65310\,{\bf k}^{10}\,\omega  - 8085\,{\bf k}^{12}\,\omega  \nonumber\\&&
- 
         358400\,{\bf k}^2\,\omega ^2 + 4218368\,{\bf k}^4\,\omega ^2 - 
         3461600\,{\bf k}^6\,\omega ^2 - 1372476\,{\bf k}^8\,\omega ^2 \nonumber\\&&
- 
         177870\,{\bf k}^{10}\,\omega ^2 + 1469440\,{\bf k}^2\,\omega ^3 - 
         9185344\,{\bf k}^4\,\omega ^3 + 1027600\,{\bf k}^6\,\omega ^3 \nonumber\\&&
+ 
         357630\,{\bf k}^8\,\omega ^3 + 48510\,{\bf k}^{10}\,\omega ^3 + 
         1075200\,\omega ^4 - 9354240\,{\bf k}^2\,\omega ^4 \nonumber\\&&
+ 
         10195136\,{\bf k}^4\,\omega ^4 + 2348976\,{\bf k}^6\,\omega ^4 + 
         404250\,{\bf k}^8\,\omega ^4 - 3978240\,\omega ^5 \nonumber\\&&
+ 
         15638560\,{\bf k}^2\,\omega ^5 - 3365600\,{\bf k}^4\,\omega ^5 - 
         777420\,{\bf k}^6\,\omega ^5 - 121275\,{\bf k}^8\,\omega ^5 \nonumber\\&&
+ 
         7869120\,\omega ^6 - 12272960\,{\bf k}^2\,\omega ^6 - 
         2002392\,{\bf k}^4\,\omega ^6 - 485100\,{\bf k}^6\,\omega ^6 \nonumber\\&&
- 
         8571360\,\omega ^7 + 3543120\,{\bf k}^2\,\omega ^7 + 
         839580\,{\bf k}^4\,\omega ^7 + 161700\,{\bf k}^6\,\omega ^7 \nonumber\\&&
+ 
         5023200\,\omega ^8 + 852600\,{\bf k}^2\,\omega ^8 + 
         323400\,{\bf k}^4\,\omega ^8 - 1231440\,\omega ^9 \nonumber\\&&
- 
         450870\,{\bf k}^2\,\omega ^9 - 121275\,{\bf k}^4\,\omega ^9 - 
         145740\,\omega ^{10} - 113190\,{\bf k}^2\,\omega ^{10} \nonumber\\&&
+ 
         96390\,\omega ^{11} + 48510\,{\bf k}^2\,\omega ^{11} + 
         16170\,\omega ^{12} - 8085\,\omega ^{13} \bigg] \bigg/ \nonumber\\
&& \bigg[ 612360\,\bigg( {\bf k}^2 + 2\,\omega  - \omega ^2 \bigg)^7\bigg] 
\end{eqnarray}
for the $3P_{3/2}$ state,
\begin{eqnarray}
{\tilde P}_{\rm ren}(3P_{3/2}) &=& 
  (Z \alpha)^2\,\bigg[ -8\,{\bf k}^2 + 4\,{\bf k}^4 + 6\,{\bf k}^2\,\omega  - 
         3\,{\bf k}^4\,\omega  + 12\,\omega ^2 - 10\,{\bf k}^2\,\omega ^2 \nonumber\\&&
- 
         6\,\omega ^3 + 6\,{\bf k}^2\,\omega ^3 + 6\,\omega ^4 - 
         3\,\omega ^5 \bigg] \bigg/ \bigg[ 
     27\,\bigg( {\bf k}^2 + 2\,\omega  - \omega ^2 \bigg) ^3\bigg] \nonumber\\&&
+ 
   (Z \alpha) ^4\,\bigg[ -128\,{\bf k}^4 + 32\,{\bf k}^6 + 8\,{\bf k}^8 + 
         32\,{\bf k}^4\,\omega  - 26\,{\bf k}^6\,\omega  - 3\,{\bf k}^8\,\omega  \nonumber\\&&
+ 
         192\,{\bf k}^2\,\omega ^2 - 68\,{\bf k}^4\,\omega ^2 - 
         30\,{\bf k}^6\,\omega ^2 - 56\,{\bf k}^2\,\omega ^3 + 
         42\,{\bf k}^4\,\omega ^3 \nonumber\\&&
+ 12\,{\bf k}^6\,\omega ^3 - 
         112\,\omega ^4 + 42\,{\bf k}^4\,\omega ^4 + 24\,\omega ^5 - 
         6\,{\bf k}^2\,\omega ^5 - 18\,{\bf k}^4\,\omega ^5 \nonumber\\&&
+ 
         36\,\omega ^6 - 26\,{\bf k}^2\,\omega ^6 - 10\,\omega ^7 + 
         12\,{\bf k}^2\,\omega ^7 + 6\,\omega ^8 - 3\,\omega ^9
          \bigg] \bigg/ \nonumber\\&&
\bigg[ 
     324\,\bigg[ {\bf k}^2 + 2\,\omega  - \omega ^2 \bigg) ^5\bigg] \nonumber\\&&
+ 
   (Z \alpha) ^6\,\bigg[ -4179968\,{\bf k}^6 + 516608\,{\bf k}^8 + 
         224616\,{\bf k}^{10} + 12180\,{\bf k}^{12} \nonumber\\&&
+ 2351104\,{\bf k}^4\,\omega  + 
         243712\,{\bf k}^6\,\omega  - 1022336\,{\bf k}^8\,\omega  - 
         103110\,{\bf k}^{10}\,\omega  \nonumber\\&&
- 3045\,{\bf k}^{12}\,\omega  - 
         1433600\,{\bf k}^2\,\omega ^2 + 8960000\,{\bf k}^4\,\omega ^2 + 
         667456\,{\bf k}^6\,\omega ^2 \nonumber\\&&
- 539868\,{\bf k}^8\,\omega ^2 - 
         66990\,{\bf k}^{10}\,\omega ^2 + 4157440\,{\bf k}^2\,\omega ^3 - 
         2845696\,{\bf k}^4\,\omega ^3 \nonumber\\&&
+ 1703632\,{\bf k}^6\,\omega ^3 + 
         370230\,{\bf k}^8\,\omega ^3 + 18270\,{\bf k}^{10}\,\omega ^3 + 
         4300800\,\omega ^4 \nonumber\\&&
- 17297280\,{\bf k}^2\,\omega ^4 - 
         3980704\,{\bf k}^4\,\omega ^4 + 135408\,{\bf k}^6\,\omega ^4 + 
         152250\,{\bf k}^8\,\omega ^4 \nonumber\\&&
- 10752000\,\omega ^5 + 
         10612000\,{\bf k}^2\,\omega ^5 - 619136\,{\bf k}^4\,\omega ^5 - 
         449820\,{\bf k}^6\,\omega ^5 \nonumber\\&&
- 45675\,{\bf k}^8\,\omega ^5 + 
         15408960\,\omega ^6 - 99680\,{\bf k}^2\,\omega ^6 + 
         586824\,{\bf k}^4\,\omega ^6 \nonumber\\&&
- 182700\,{\bf k}^6\,\omega ^6 - 
         10647840\,\omega ^7 + 216720\,{\bf k}^2\,\omega ^7 + 
         159180\,{\bf k}^4\,\omega ^7 \nonumber\\&&
+ 60900\,{\bf k}^6\,\omega ^7 + 
         2896320\,\omega ^8 - 543480\,{\bf k}^2\,\omega ^8 + 
         121800\,{\bf k}^4\,\omega ^8 \nonumber\\&&
- 278880\,\omega ^9 + 
         65730\,{\bf k}^2\,\omega ^9 - 45675\,{\bf k}^4\,\omega ^9 + 
         136500\,\omega ^{10} \nonumber\\&&
- 42630\,{\bf k}^2\,\omega ^{10} - 
         42210\,\omega ^{11} + 18270\,{\bf k}^2\,\omega ^{11} + 
         6090\,\omega ^{12} \nonumber\\
&& - 3045\,\omega ^{13} \bigg] \bigg/ \bigg[ 
2449440\,\bigg( {\bf k}^2 + 2\,\omega  - \omega ^2 \bigg) ^7\bigg]
\end{eqnarray}
for the $4P_{1/2}$ state,
\begin{eqnarray}
{\tilde P}_{\rm ren}(4P_{1/2}) &=& 
  (Z \alpha) ^2\,\bigg[ -8\,{\bf k}^2 + 4\,{\bf k}^4 + 6\,{\bf k}^2\,\omega  - 
         3\,{\bf k}^4\,\omega  + 12\,\omega ^2 - 10\,{\bf k}^2\,\omega ^2 \nonumber\\&&
- 
         6\,\omega ^3 + 6\,{\bf k}^2\,\omega ^3 + 6\,\omega ^4 - 
         3\,\omega ^5 \bigg] \bigg/ \bigg[ 
 48\,\bigg( {\bf k}^2 + 2\,\omega  - \omega ^2 \bigg) ^3\bigg] \nonumber\\&&
+ 
   (Z \alpha) ^4\,\bigg[ -2944\,{\bf k}^4 + 1072\,{\bf k}^6 + 520\,{\bf k}^8 + 
         2080\,{\bf k}^4\,\omega  - 10\,{\bf k}^6\,\omega  \nonumber\\&&
- 195\,{\bf k}^8\,\omega  + 
         5760\,{\bf k}^2\,\omega ^2 - 3412\,{\bf k}^4\,\omega ^2 - 
         1950\,{\bf k}^6\,\omega ^2 - 3640\,{\bf k}^2\,\omega ^3 \nonumber\\&&
+ 
         1050\,{\bf k}^4\,\omega ^3 + 780\,{\bf k}^6\,\omega ^3 - 
         560\,\omega ^4 + 5040\,{\bf k}^2\,\omega ^4 + 
         2730\,{\bf k}^4\,\omega ^4 \nonumber\\&&
+ 1560\,\omega ^5 - 
         2070\,{\bf k}^2\,\omega ^5 - 1170\,{\bf k}^4\,\omega ^5 - 
         2700\,\omega ^6 - 1690\,{\bf k}^2\,\omega ^6 \nonumber\\&&
+ 
         1030\,\omega ^7 + 780\,{\bf k}^2\,\omega ^7 + 
         390\,\omega ^8 - 195\,\omega ^9 \bigg] \bigg/ \nonumber\\&& \bigg[ 
15360\,\bigg( {\bf k}^2 + 2\,\omega  - \omega ^2 \bigg) ^5\bigg] \nonumber\\&&
+ 
   (Z \alpha) ^6\,\bigg[ -41518080\,{\bf k}^6 + 20285056\,{\bf k}^8 + 
         12019560\,{\bf k}^{10} \nonumber\\&&
+ 1140300\,{\bf k}^{12} + 21790720\,{\bf k}^4\,\omega  + 
         67517184\,{\bf k}^6\,\omega  - 1295056\,{\bf k}^8\,\omega  \nonumber\\&&
- 
         2875250\,{\bf k}^{10}\,\omega  - 285075\,{\bf k}^{12}\,\omega  - 
         13189120\,{\bf k}^2\,\omega ^2 + 164921344\,{\bf k}^4\,\omega ^2 \nonumber\\&&
- 
         138538016\,{\bf k}^6\,\omega ^2 - 50829380\,{\bf k}^8\,\omega ^2 - 
         6271650\,{\bf k}^{10}\,\omega ^2 \nonumber\\&&
+ 53975040\,{\bf k}^2\,\omega ^3 - 
         357889728\,{\bf k}^4\,\omega ^3 + 50610672\,{\bf k}^6\,\omega ^3 \nonumber\\&&
+ 
         14327250\,{\bf k}^8\,\omega ^3 + 1710450\,{\bf k}^{10}\,\omega ^3 + 
         39567360\,\omega ^4 - 357683200\,{\bf k}^2\,\omega ^4 \nonumber\\&&
+ 
         404953920\,{\bf k}^4\,\omega ^4 + 81291280\,{\bf k}^6\,\omega ^4 + 
         14253750\,{\bf k}^8\,\omega ^4 \nonumber\\&&
- 147302400\,\omega ^5 + 
601024480\,{\bf k}^2\,\omega ^5 - 153435296\,{\bf k}^4\,\omega ^5 \nonumber\\&&
- 
         28556500\,{\bf k}^6\,\omega ^5 - 4276125\,{\bf k}^8\,\omega ^5 + 
         294387520\,\omega ^6 - 485016000\,{\bf k}^2\,\omega ^6 \nonumber\\&&
- 
         59093160\,{\bf k}^4\,\omega ^6 - 17104500\,{\bf k}^6\,\omega ^6 - 
         325403680\,\omega ^7 + 160218800\,{\bf k}^2\,\omega ^7 \nonumber\\&&
+ 
         28458500\,{\bf k}^4\,\omega ^7 + 5701500\,{\bf k}^6\,\omega ^7 + 
         198315040\,\omega ^8 + 17532200\,{\bf k}^2\,\omega ^8 \nonumber\\&&
+ 
         11403000\,{\bf k}^4\,\omega ^8 - 56099120\,\omega ^9 - 
         14180250\,{\bf k}^2\,\omega ^9 - 4276125\,{\bf k}^4\,\omega ^9 \nonumber\\&&
- 
         920500\,\omega ^{10} - 3991050\,{\bf k}^2\,\omega ^{10} + 
         2826250\,\omega ^{11} + 1710450\,{\bf k}^2\,\omega ^{11} \nonumber\\&&
+ 
         570150\,\omega ^{12} - 285075\,\omega ^{13} \bigg] \bigg/ \bigg[ 
     51609600\,\bigg( {\bf k}^2 + 2\,\omega  - \omega ^2 \bigg)^7\bigg] 
\end{eqnarray}
and for the $4P_{3/2}$ state
\begin{eqnarray}
{\tilde P}_{\rm ren}(4P_{3/2}) &=& 
  (Z \alpha) ^2\,\bigg[ -8\,{\bf k}^2 + 4\,{\bf k}^4 + 6\,{\bf k}^2\,\omega  - 
         3\,{\bf k}^4\,\omega  + 12\,\omega ^2 - 10\,{\bf k}^2\,\omega ^2 \nonumber\\&&
- 
         6\,\omega ^3 + 6\,{\bf k}^2\,\omega ^3 + 6\,\omega ^4 - 
         3\,\omega ^5 \bigg] \bigg/ \bigg[ 
     48\,\bigg( {\bf k}^2 + 2\,\omega  - \omega ^2 \bigg) ^3\bigg] \nonumber\\&&
+ 
   (Z \alpha) ^4\,\bigg[ -2944\,{\bf k}^4 + 752\,{\bf k}^6 + 200\,{\bf k}^8 + 
         800\,{\bf k}^4\,\omega  - 570\,{\bf k}^6\,\omega  \nonumber\\&&
- 75\,{\bf k}^8\,\omega  + 
         4480\,{\bf k}^2\,\omega ^2 - 1652\,{\bf k}^4\,\omega ^2 - 
         750\,{\bf k}^6\,\omega ^2 - 1400\,{\bf k}^2\,\omega ^3 \nonumber\\&&
+ 
         970\,{\bf k}^4\,\omega ^3 + 300\,{\bf k}^6\,\omega ^3 - 
         2480\,\omega ^4 + 240\,{\bf k}^2\,\omega ^4 + 
         1050\,{\bf k}^4\,\omega ^4 \nonumber\\&&
+ 600\,\omega ^5 - 
         230\,{\bf k}^2\,\omega ^5 - 450\,{\bf k}^4\,\omega ^5 + 
         660\,\omega ^6 - 650\,{\bf k}^2\,\omega ^6 \nonumber\\&&
- 
         170\,\omega ^7 + 300\,{\bf k}^2\,\omega ^7 + 
         150\,\omega ^8 - 75\,\omega ^9 \bigg] \bigg/ \nonumber\\&&
\bigg[ 15360\,\bigg( {\bf k}^2 + 2\,\omega 
- \omega ^2 \bigg) ^5\bigg] \nonumber\\&&
+ 
   (Z \alpha) ^6\,\bigg[ -41518080\,{\bf k}^6 + 5273472\,{\bf k}^8 + 
         2372328\,{\bf k}^{10} + 132300\,{\bf k}^{12} \nonumber\\&&
+ 21790720\,{\bf k}^4\,\omega  + 
         3438848\,{\bf k}^6\,\omega  - 10087504\,{\bf k}^8\,\omega  - 
         1069810\,{\bf k}^{10}\,\omega  \nonumber\\&&
- 33075\,{\bf k}^{12}\,\omega  - 
         13189120\,{\bf k}^2\,\omega ^2 + 88747008\,{\bf k}^4\,\omega ^2 + 
         4454624\,{\bf k}^6\,\omega ^2 \nonumber\\&&
- 5916484\,{\bf k}^8\,\omega ^2 - 
         727650\,{\bf k}^{10}\,\omega ^2 + 37847040\,{\bf k}^2\,\omega ^3 - 
         29007552\,{\bf k}^4\,\omega ^3 \nonumber\\&&
+ 18082288\,{\bf k}^6\,\omega ^3 + 
         3870930\,{\bf k}^8\,\omega ^3 + 198450\,{\bf k}^{10}\,\omega ^3 + 
         39567360\,\omega ^4 \nonumber\\&&
- 163681280\,{\bf k}^2\,\omega ^4 - 
         34772416\,{\bf k}^4\,\omega ^4 + 2115344\,{\bf k}^6\,\omega ^4 \nonumber\\&&
+ 
         1653750\,{\bf k}^8\,\omega ^4 - 98918400\,\omega ^5 + 
         100285920\,{\bf k}^2\,\omega ^5 - 8639904\,{\bf k}^4\,\omega ^5 \nonumber\\&&
- 
         4785620\,{\bf k}^6\,\omega ^5 - 496125\,{\bf k}^8\,\omega ^5 + 
         143286080\,\omega ^6 - 1740480\,{\bf k}^2\,\omega ^6 \nonumber\\&&
+ 
         5429592\,{\bf k}^4\,\omega ^6 - 1984500\,{\bf k}^6\,\omega ^6 - 
         99145760\,\omega ^7 + 3382960\,{\bf k}^2\,\omega ^7 \nonumber\\&&
+ 
         1829380\,{\bf k}^4\,\omega ^7 + 661500\,{\bf k}^6\,\omega ^7 + 
         26784800\,\omega ^8 - 5400920\,{\bf k}^2\,\omega ^8 \nonumber\\&&
+ 
         1323000\,{\bf k}^4\,\omega ^8 - 2737840\,\omega ^9 + 
         563430\,{\bf k}^2\,\omega ^9 - 496125\,{\bf k}^4\,\omega ^9 \nonumber\\&&
+ 
         1400140\,\omega ^{10} - 463050\,{\bf k}^2\,\omega ^{10} - 
         408310\,\omega ^{11} + 198450\,{\bf k}^2\,\omega ^{11} \nonumber\\&&
+ 
         66150\,\omega ^{12} - 33075\,\omega ^{13} \bigg] \bigg/ \bigg[ 
     51609600\,\bigg( {\bf k}^2 + 2\,\omega  - \omega ^2 \bigg)^7\bigg]
\end{eqnarray}
Having calculated $\tilde P$, we finally integrate along $C_H$ to obtain the
result for $E_H$,
\begin{equation}
\label{getEH}
E_H = - i e^2 m \, \int_{C_H} \frac{d\omega}{2 \pi} 
\int \frac{d^3 {\bf k}}{(2 \pi)^3} \,
\left[ \frac{1}{\omega^2 - {\bf k}^2} - 
  \frac{1}{\omega^2 - {\bf k}^2 - M^2} \right]\,{\tilde P}_{\rm ren}.
\end{equation}
Note that as ${\tilde P}_{\rm ren} = {\tilde P} - \delta m_l$,
both $\tilde P$ as well as the local mass renormalization term $\delta m_l$
are properly integrated with the regularized photon propagator.

The final integrations with respect to the photon momenta are done in a different way 
for the terms in  ${\tilde P}_{\rm ren}$  which require regularization and 
those which do not. The terms which require regularization are integrated 
covariantly by Feynman parameter techniques and a subsequent Wick rotation.
These terms are not integrated along $C_H$, but rather along $C'_H$.
They are not infrared divergent, so we may put $\epsilon = 0$ 
for these terms. 
Those terms which do not require regularization are integrated in an 
essentially 
non-covariant way. The $d^3 {\bf k}$ integration is carried out first, then
we proceed to the $d \omega$ integration. 

The integration procedure for the terms which require regularization is
as follows. We isolate in $(\tilde P - \delta m_l)$ those terms which 
would be ultraviolet divergent if integrated with the unregularized 
photon propagator. We denote these terms by ${\tilde P}_{\rm ren}^{\rm div}$. 
We then evaluate
\begin{displaymath}
\delta E_{\rm div} = \int_{C_H} \frac{d\omega}{2 \pi} 
\int \frac{d^3 {\bf k}}{(2 \pi)^3} \frac{1}{\omega^2 - {\bf k}^2} \,
\left[ {\tilde P}_{\rm ren}^{\rm div} \right].
\end{displaymath}
All terms in ${\tilde P}_{\rm ren}^{\rm div}$
require regularization. 

By simple power counting arguments,
it can be shown that  ${\tilde P}_{\rm ren}^{\rm div}$ 
is exactly the sum of those terms
of order $k^{n}$
where $n \geq -2$ for large $k$. Therefore, the terms contributing to 
${\tilde P}_{\rm ren}^{\rm div}$ can easily be isolated.
The terms in ${\tilde P}_{\rm div}$, can obviously 
be written as the sum of terms of the form
\begin{equation}
\label{pdiv}
{\tilde P}_{\rm div} = 
\sum_i \frac{p_i(|{\bf k}|,\omega)}{\left[{\bf k}^2 + 2 \omega - \omega^2\right]^{n_i}}
\end{equation}
where $p_i$ is a polynomial in $|{\bf k}|$ and $\omega$
and ${\rm deg}(p_i) \geq n_i - 2$. The entirely covariant 
integration procedure for the divergent terms will be outlined here. 
The terms in Eq. (\ref{pdiv}) need to be multiplied by the factors 
$1/(\omega^2 - {\bf k}^2)$ and $1/(\omega^2 - {\bf k}^2 -M^2)$
from the regularized photon propagator. We use Feynman parameters
in the form
\begin{equation}
\label{defx}
\frac{1}{A\,B^n}=\int_0^1 dx\,\frac{n x^{n-1}}{\left[A\,(1-x)+ B x\right]^n}
\end{equation}
to join the denominators. Identifying $A_1 = - \omega^2 + {\bf k}^2$,
$A_2 = - \omega^2 + {\bf k}^2 + M^2$ and
$B = {\bf k}^2 + 2 \omega - \omega^2$, we have for the contribution from the
unrenormalized photon propagator $1/(\omega^2 - {\bf k}^2)$,
\begin{equation}
A_1\,(1-x)+ B x = - {\tilde k}^2 + D_1
\end{equation}
where
\begin{equation}
{\tilde k} = ({\tilde \omega}, {\bf k}) = (\omega - x, {\bf k}) 
\quad \mbox{and} \quad D_1 = x^2,
\end{equation}
and for the contribution from the renormalization part of the photon propagator
\begin{equation}
A\,(1-x)+ B x = - {\tilde k}^2 + D_2 
\end{equation}
where
\begin{equation}
D_2 = x^2 + M\,(1-x).
\end{equation}
The energy shift due to the divergent terms is then proportional to
\begin{displaymath}
\sum_i \int_0^1 dx \, \int d{\tilde \omega} \, d^3 k \, 
\frac{n_i \, x^{n_i-1} \, p_i(|{\bf k}|,{\tilde \omega}) }{- {\tilde k}^2 + D},
\end{displaymath}
where $D$ represents either of the terms $D_1$ or $D_2$ 
and it is understood that
the contribution from these two terms must be subtracted to obtain the final
result. Performing the Wick rotation
\begin{displaymath}
{\tilde \omega} \to i \omega,
\end{displaymath}
we have
\begin{displaymath}
- {\tilde k}^2 = \omega^2 + {\bf k}^2 = k_e^2 + D,
\end{displaymath}
where $k_e$ is the Euclidean 4-vector $k_e = (\omega, {\bf k})$. 
We then obtain the
energy shift due to the divergent terms proportional to
\begin{displaymath}
\sum_i \int_0^1 dx \, \int d^4 k_e 
\frac{n_i \, x^{n_i-1} \, p_i(|{\bf k}|,\omega) }{k_e^2 + D}.
\end{displaymath}
The (straightforward) angular part of these integrals can 
be done by parametrizing the 
Euclidean 4-space as $\omega = k_e\, \cos\gamma$,
$k^1 = k_e \, \sin\gamma \, \sin\theta \, \cos\phi$,
$k^2 = k_e \, \sin\gamma \, \sin\theta \, \sin\phi$,
$k_3 = k_e \, \sin\gamma \, \cos\theta$. We then have 
$\omega =  k_e\, \cos\gamma$, $|{\bf k}| =  k_e \, \sin\gamma$. For the average
over the 4-dimensional angle, we utilize the formulae
\begin{displaymath}
\int \frac{d \Omega_e^{(4)}}{2 \pi^2} \,\cos^2\gamma = \frac{1}{4}, 
\quad 
\int \frac{d \Omega_e^{(4)}}{2 \pi^2} \,\cos^4\gamma = \frac{1}{8}, 
\quad 
\int \frac{d \Omega_e^{(4)}}{2 \pi^2} \,\cos^6\gamma = \frac{5}{64}. 
\end{displaymath}
The remaining radial part of the integrals can be evaluated with 
the help of the formulae
\begin{displaymath}
\int_0^{\infty} dk_e \, k_e^3 \, \frac{1}{(k_e^2 + D)^{\beta}} = 
\frac{1}{2 D^{\beta-2} \, (\beta-1)\,(\beta-2)} \quad (\beta > 2)
\end{displaymath}
and
\begin{displaymath}
\int_0^{\Lambda} dk_e \,k_e^3 \,\frac{1}{(k_e^2 + D)^2} = 
\frac{1}{2}\, \ln\left(\frac{\Lambda}{D}\right) + 
O\left(\frac{1}{\Lambda}\right)^2.
\end{displaymath}
It is easy to prove that the dependence on the temporary upper cutoff
$\Lambda$ disappears when the contribution
due to the unregularized and the regularization part of the 
photon propagator are subtracted ($D_1$ and $D_2$). 

As the final step, we integrate over the 
parameter $x$ introduced in Eq. (\ref{defx}) and subsequently investigate 
the resulting expression in the limit $M \to \infty$. The respective 
expressions vanish as $M \to \infty$ for all states considered in this
paper. This fact is intimately linked to our
using the entirely covariant Feynman parameter 
approach for the integration of the
divergent terms. If we had used a non-covariant scheme of integration, as in 
\cite{krp1}, then we would have had to take into account finite correction
terms to obtain the correct result for the Lamb shift.

Note that the (divergent) spurious terms of 
order $(Z \alpha)^2$, which are present in all
of the matrix elements ${\tilde P}_{\rm ren}$, 
vanish after we have performed the
$d^4 k$ integration in the way outlined above, which includes
final expansion in the $\epsilon$ parameter.

The terms in ${\tilde P}_{\rm ren}^{\rm fin}$ which are finite 
when integrated with
the unrenormalized photon propagator, do not need regularization. 
It is easy to see that ${\tilde P}_{\rm ren}^{\rm fin}$ can be written
as the sum of terms of the form
\begin{equation}
\label{pfin}
{\tilde P}_{\rm ren}^{\rm fin} = 
\sum_j 
\frac{q_j(|{\bf k}|,\omega)}
{\left[{\bf k}^2 + 2 \omega - \omega^2\right]^{n_j}},
\end{equation}
where $q_j$ is a polynomial in $|{\bf k}|$ and $\omega$ whose degree 
is less than $2 n_j - 2$ to insure ultraviolet convergence. For the 
$4P_{1/2}$ state, e. g., ${\tilde P}_{\rm ren}^{\rm fin}$ is given by
\begin{eqnarray}
P_{\rm ren}^{\rm fin} &=&
  (Z \alpha)^2\,\bigg[ -4\,{\bf k}^2 + 3\,{\bf k}^2\,\omega  + 6\,\omega ^2 - 
        3\,\omega ^3 \bigg] \bigg/ \bigg[ 
    24\,\bigg( {\bf k}^2 + 2\,\omega  - \omega ^2 \bigg) ^3\bigg] \nonumber\\&&
+ (Z \alpha) ^4\,\bigg[ -1472\,{\bf k}^4 + 
        536\,{\bf k}^6 + 1040\,{\bf k}^4\,\omega  - 
        5\,{\bf k}^6\,\omega  + 2880\,{\bf k}^2\,\omega ^2 \nonumber\\&&
        - 1706\,{\bf k}^4\,\omega ^2 - 1820\,{\bf k}^2\,\omega ^3 + 
        525\,{\bf k}^4\,\omega ^3 - 280\,\omega ^4 + 
        2520\,{\bf k}^2\,\omega ^4 \nonumber\\&&
        + 780\,\omega ^5 - 1035\,{\bf k}^2\,\omega ^5 - 1350\,\omega ^6 + 515\,\omega ^7
         \bigg] \bigg/ \nonumber\\&& 
    \bigg[ 7680\,\bigg( {\bf k}^2 + 2\,\omega  - \omega ^2 \bigg)^5 \bigg] \nonumber\\&&
+ (Z \alpha) ^6\,\bigg[ -20759040\,{\bf k}^6 + 10142528\,{\bf k}^8 + 
        6009780\,{\bf k}^{10} \nonumber\\&&
        + 10895360\,{\bf k}^4\,\omega  + 
        33758592\,{\bf k}^6\,\omega  - 647528\,{\bf k}^8\,\omega  - 
        1437625\,{\bf k}^{10}\,\omega  \nonumber\\&&
        - 6594560\,{\bf k}^2\,\omega ^2 + 
        82460672\,{\bf k}^4\,\omega ^2 - 69269008\,{\bf k}^6\,\omega ^2 - 
        25414690\,{\bf k}^8\,\omega ^2 \nonumber\\&&
        + 26987520\,{\bf k}^2\,\omega ^3 - 
        178944864\,{\bf k}^4\,\omega ^3 + 25305336\,{\bf k}^6\,\omega ^3 \nonumber\\&&
        + 7163625\,{\bf k}^8\,\omega ^3 + 19783680\,\omega ^4 - 
        178841600\,{\bf k}^2\,\omega ^4 + 202476960\,{\bf k}^4\,\omega ^4 \nonumber\\&&
        + 40645640\,{\bf k}^6\,\omega ^4 - 73651200\,\omega ^5 + 
        300512240\,{\bf k}^2\,\omega ^5 - 76717648\,{\bf k}^4\,\omega ^5 \nonumber\\&&
        - 14278250\,{\bf k}^6\,\omega ^5 + 147193760\,\omega ^6 - 
        242508000\,{\bf k}^2\,\omega ^6 - 29546580\,{\bf k}^4\,\omega ^6 \nonumber\\&&
        - 162701840\,\omega ^7 + 80109400\,{\bf k}^2\,\omega ^7 + 
        14229250\,{\bf k}^4\,\omega ^7 + 99157520\,\omega ^8 \nonumber\\&&
        + 8766100\,{\bf k}^2\,\omega ^8 - 28049560\,\omega ^9 - 
        7090125\,{\bf k}^2\,\omega ^9 - 460250\,\omega ^{10} \nonumber\\&&
        + 1413125\,\omega ^{11} \bigg] \bigg/ \bigg[ 
    25804800\,\bigg( {\bf k}^2 + 2\,\omega  - \omega ^2 \bigg) ^7\bigg]
\end{eqnarray}
We have to calculate
\begin{displaymath}
\delta E_H = - i e^2 m \, \int_{C_H} \frac{d\omega}{2 \pi} 
\int \frac{d^3 {\bf k}}{(2 \pi)^3} \frac{1}{\omega^2 - {\bf k}^2} \,
\left[ {\tilde P}_{\rm ren}^{\rm fin} \right].
\end{displaymath}
The quantity $F_H$ defined by
\begin{equation}
E_H = \frac{\alpha}{\pi}\,m\,\frac{(Z \alpha)^4}{n^3} \, F_H
\end{equation}
in Eq. (\ref{defF}) may be expressed as
\begin{equation}
F_H = n^3/(Z \alpha)^4 \, \int_{C_H} d\omega \, {\cal F}(\omega)
\end{equation}
where
\begin{displaymath}
{\cal F}(\omega) = \frac{1}{2 \pi i} \,
\int d|{\bf k}| \, \frac{{\bf k}^2}{\omega^2 - {\bf k}^2} \, 
{\tilde P}_{\rm ren}^{\rm fin}(\omega, |{\bf k}|). 
\end{displaymath}
One can dispose of the factors ${\bf k}^2$ in 
the numerator of the integrand using the following procedure. 
First write ${\bf k}^2$ as
\begin{displaymath}
{\bf k}^2 = Y - 2 \omega + \omega^2,
\end{displaymath}
so $Y = {\bf k}^2 + 2 \omega - \omega^2$ corresponds to the denominator
in Eq. (\ref{pfin}). The resulting expression is subsequently expanded. 
The powers of $Y$ cancel,
and the result does not carry powers of ${\bf k}$ in the numerator
any more. 


The integrand in ${\cal F}(\omega)$ can be written as the sum of 
terms of the form
\begin{equation}
\label{intk}
{\cal F}(\omega) = \frac{1}{2 \pi i} \,
\int d|{\bf k}| \, \left( \sum_j a_j 
\frac{\omega^{n_j}}{\omega^2 - {\bf k}^2} \,
\frac{1}{\left({\bf k}^2 + 2 \omega - \omega^2\right)^{m_j}} \right)
\end{equation}
with suitable coefficients $a_j$. The $d|{\bf k}|$ integration can then be 
carried out using the formula 
\begin{eqnarray}
\label{intformula}
&&\vspace{-4 ex} \frac{1}{2 \pi i} \int_{-\infty}^{\infty} d|{\bf k}| \, 
\frac{1}{\omega^2 - {\bf k}^2} \, \frac{1}{{\bf k}^2 + \Omega} \nonumber\\
&& = \frac{-1}{2 \, {\rm sgn}({\rm Im}(\omega)) \, 
(\omega^2 + \Omega) \, \omega} \, 
+ \frac{(-1)^{n-1}}{(n-1)!} \, \frac{\partial^{n-1}}{\partial \Omega^{n-1}}\,
\left[ \frac{i}{2\,\sqrt{\Omega}}\,\frac{1}{\omega^2 + \Omega} \right]
\end{eqnarray}
where ${\rm sgn}$ is the sign function defined as 
\begin{eqnarray}
{\rm sgn}(x) = \Biggl\{
\begin{array}{rr}
1 & x \geq 0\\
-1 & x <0
\end{array} 
\end{eqnarray}
We identify
\begin{equation}
\Omega = 2 \omega - \omega^2
\end{equation}
in order to carry out the integration in Eq. (\ref{intk}).
The formula Eq. (\ref{intformula}) deserves some comments. 
We are integrating along 
the real axis. Because $\omega$ and $\Omega$
both have an infinitesimal imaginary part all along the
contour of integration $C_H$, the positions of the poles of the 
integrand are well defined. 

The branch cuts of the function ${\cal F}(\omega)$ can be readily identified.
Due to the term ${\rm sgn}({\rm Im}(\omega)) \, \omega$ in the first term on the
right hand side in (\ref{intformula}), there is a branch cut along the 
positive real axis. This branch cut is caused by the photon propagator.
Due to the term $\sqrt{\Omega}$ in the second term on the
right hand side in (\ref{intformula}) there is also
a branch cut along the line where the expression
$\Omega = 2 \omega - \omega^2$ assumes negative real values. This branch cut
extends from $\omega = 2$ along the positive real axis to $\omega = \infty$.
It is caused by the Dirac-Coulomb propagator (see Fig. \ref{intcontour1}).

It can be explicitly checked that the function ${\cal F}(\omega)$ satisfies the
equation
\begin{equation}
\label{propertiesF}
{\cal F}(\omega^*) = - {\cal F}(\omega)^*.
\end{equation}   
We can divide the contour $C_H$ in an upper contour $C_H^u$ which extends from
$\epsilon + i 0^{+}$ to $\infty + i 0^{+}$ and a lower contour 
$C_H^l$ which extends from $\infty - i 0^{+}$ to $\epsilon - i 0^{+}$. We
then have due to Eq. (\ref{propertiesF})
\begin{eqnarray}
\int_{C_H} d\omega \,{\cal F}(\omega) &=&
\int_{C_H^l}  d\omega \,{\cal F}(\omega) +
\int_{C_H^u}  d\omega \,{\cal F}(\omega) \nonumber\\
&=& \int_{\epsilon}^{\infty}  d\omega \,{\cal F}(\omega + i 0^{+}) +
\int_{\infty}^{\epsilon}  d\omega \,{\cal F}(\omega - i 0^{+}) \nonumber\\
&=& \int_{\epsilon}^{\infty}  d\omega \,{\cal F}(\omega + i 0^{+}) +
\int_{\epsilon}^{\infty}  d\omega \,{\cal F}(\omega + i 0^{+})^* \nonumber\\
&=& \int_{C_H^u}  d\omega \,{\cal F}(\omega) + c.c.
\end{eqnarray}
We restrict ourselves therefore to the upper contour $C_H^u$ and
understand that the complete result is the sum of the integral along the 
upper contour plus its complex conjugate. 

We then perform a change of variable to proceed to the final
$d\omega$ integration. Defining
\begin{equation}
u = \frac{\sqrt{\Omega} + i \omega}{\sqrt{\Omega} - i \omega}.
\end{equation}
and
\begin{equation}
{\cal U}(u) = \frac{d\omega}{du} \, {\cal F}(\omega),
\end{equation}
we have
\begin{equation}
F_H = n^3/(Z \alpha)^4 \, \int_{u(C_H)} du \, {\cal U}(u).
\end{equation}
Note that
\begin{equation}
u = \frac{\sqrt{{\rm Re}(2\,\omega - \omega^2)} + i \, \omega}
{\sqrt{{\rm Re}(2\,\omega - \omega^2)} - i \, \omega} \quad \mbox{for} \quad
\omega \in C_H^U, {\rm Re}(\omega) \in [\epsilon, 2[,
\end{equation}
whereas
\begin{equation}
u = \frac{\sqrt{{\rm Re}(\omega^2 - 2 \, \omega)} - \, \omega}
{\sqrt{{\rm Re}(\omega^2 - 2 \, \omega)} + \, \omega} \quad \mbox{for} \quad
\omega \in C_H^U, {\rm Re}(\omega) \in [2, \infty),
\end{equation}
where the argument of the square root is written in such a way as to
represent a positive real quantity in the two above equations.
So $u(\omega = 0) = 1$, $u(\omega = 2) = -1$, $u(\infty + i\,0^{+}) = 0$,
and $u \in [-1, 0)$ for $\omega \in C_H^U, {\rm Re}(\omega) \in [2, \infty)$. 
So for $\omega \in C_H^U$, 
${\rm Re}(\omega) \in [\epsilon, 2)$, $u$ has a nonvanishing 
imaginary part, whereas for 
$\omega \in C_H^U$, ${\rm Re}(\omega) \in [2, \infty)$, $u$ is a real quantity.
The mapping $\omega \to u$ is one-on-one for $\omega \in C_H^U$.

Note that if we had chosen the lower contour $C_H^L$, then 
$u(\infty - i\,0^{+}) = -\infty$. In that case, the above substitution
would not have had the desired property $u \to 0$ for $\omega \to \infty$.

On $C_H^U$, $\omega$ and $\sqrt{\Omega}$ can be expressed as functions of $u$
according to
\begin{eqnarray}
\omega &=& -\frac{1}{2}\, \frac{1-u^2}{u}, \nonumber\\
\sqrt{\Omega} &=& -\frac{i}{2}\,\frac{(1-u)(1+u)}{(-u)}, \nonumber
\end{eqnarray}
i. e. $\sqrt{\Omega}$ extends along the negative imaginary axis for 
${\rm Re}(\omega) > 2$, $u \in [-1, 0]$. The result for 
${\cal U}(u)$ ($4P_{1/2}$ state) is
\begin{eqnarray}
\label{resFu}
{\cal U}(u) &=& ({Z \alpha })^2\,\bigg[ -{1\over {192}} - 
     {1\over {96\,{{\left( -1 + u \right) }^2}}} \bigg] \nonumber\\&& 
+ {({Z \alpha })^4}\,\bigg[ -{{47}\over {20480}} + 
     {{23}\over {10240\,{{\left( -1 + u \right) }^4}}} + 
     {{23}\over {10240\,{{\left( -1 + u \right) }^3}}} \nonumber\\&&+ 
     {{79}\over {40960\,{{\left( -1 + u \right) }^2}}} + 
     {{13}\over {2048\,{{\left( 1 + u \right) }^4}}} - 
     {{13}\over {2048\,{{\left( 1 + u \right) }^3}}} + 
     {{113}\over {24576\,{{\left( 1 + u \right) }^2}}} \bigg]  \nonumber\\&&
+ {({Z \alpha })^6}\,\bigg[ -{{141737}\over {41287680}} - 
     {{39173}\over {20643840\,{{\left( -1 + u \right) }^6}}} - 
     {{39173}\over {10321920\,{{\left( -1 + u \right) }^5}}} \nonumber\\&&- 
     {{2228617}\over {206438400\,{{\left( -1 + u \right) }^4}}} - 
     {{3485527}\over {206438400\,{{\left( -1 + u \right) }^3}}} - 
     {{4685519}\over {330301440\,{{\left( -1 + u \right) }^2}}} \nonumber\\&&- 
     {{499}\over {46080\,\left( -1 + u \right) }} - 
     {{343\,u}\over {614400}} + 
     {{1267}\over {65536\,{{\left( 1 + u \right) }^8}}} \nonumber\\&&- 
     {{3801}\over {65536\,{{\left( 1 + u \right) }^7}}} + 
     {{11765}\over {131072\,{{\left( 1 + u \right) }^6}}} - 
     {{2715}\over {32768\,{{\left( 1 + u \right) }^5}}} \nonumber\\&& + 
     {{69651}\over {1310720\,{{\left( 1 + u \right) }^4}}} - 
     {{28021}\over {1310720\,{{\left( 1 + u \right) }^3}}} + 
     {{89779}\over {15728640\,{{\left( 1 + u \right) }^2}}} \bigg] 
\end{eqnarray}
The next step in the calculation is the $u$ integration,
\begin{equation}
\label{finalU}
F_H = n^3/(Z \alpha)^4 \, \int_{u(\epsilon+ i\,0^{+})}^0 du \, {\cal U}(u)
+ c. c..
\end{equation}
where
\begin{equation}
u((\epsilon+ i\,0^{+}) = 1 + i\,\sqrt{2}\,\epsilon - \epsilon -
   \frac{i \epsilon^{3/2}}{2 \sqrt{2}} + O(\epsilon)^{5/2}.
\end{equation}
It is useful to define
\begin{equation}
{\tilde \epsilon} = 1 - u((\epsilon+ i\,0^{+}) =
  - i\,\sqrt{2 \epsilon} + \epsilon +
   \frac{i \epsilon^{3/2}}{2 \sqrt{2}} + O(\epsilon)^{5/2}.
\end{equation}
Note that the $\epsilon$-prescription calls for carrying out the 
$u$-integration from $u(\epsilon+ i\,0^{+})$ to $0$ and subsequently 
expanding the result in powers of $\epsilon$ up to $\epsilon^0$. 
For those terms which are finite when integrated
from $1$ to $0$ we may carry out this integration without regarding the
dependence on $\epsilon$. For instance,
\begin{equation}
\int_{u(\epsilon+ i\,0^{+})}^0 du \, \frac{1}{(1+u)^n} =
  \frac{1 - 2^{1-n}}{n - 1} + O(\sqrt{\epsilon}),
\end{equation}
so in the limit $\epsilon \to 0$, the $\epsilon$ dependent term vanishes.
Terms of the form $1/(1-u)^n$, however, introduce a 
divergence in $1/{\tilde \epsilon}$. We have
\begin{equation}
\int_{u(\epsilon+ i\,0^{+})}^0 du \, \frac{1}{(1-u)^n} =
\frac{{\tilde \epsilon}^{1-n}-1}{n-1}.
\end{equation}
We then add to the result of this integration the complex conjugate
and subsequently expand in powers of $\epsilon$.  This procedure 
is illustrated with some examples.
For the terms
proportional to $1/(1-u)^2$ we have
\begin{equation}
\int_{u(\epsilon+ i\,0^{+})}^0 du \, \frac{1}{(1-u)^2} =
1 - \frac{1}{{\tilde \epsilon}},
\end{equation}
so 
\begin{eqnarray}
\label{res1}
\int_{u(\epsilon+ i\,0^{+})}^0 du \, \frac{1}{(1-u)^2} + c.c. &=&
2 - \frac{1}{{\tilde \epsilon}} -\frac{1}{{\tilde \epsilon}^*} \nonumber\\
&& = 2 - \frac{1}{ - i\,\sqrt{2 \epsilon} + \epsilon + O(\epsilon)^{3/2}} - 
  \frac{1}{ i\,\sqrt{2 \epsilon} + \epsilon O(\epsilon)^{3/2}} \nonumber\\
&& = 2 - 
\left(\frac{i}{\sqrt{2 \epsilon}} + \frac{1}{2} + O(\epsilon)^{1/2}\right) -
\left(-\frac{i}{\sqrt{2 \epsilon}} + \frac{1}{2} + O(\epsilon)^{1/2}\right)
\nonumber\\
&& = 2 - 1 + O(\epsilon)^{1/2} = 1 + O(\epsilon)^{1/2}.  
\end{eqnarray}
For terms proportional to $1/(1-u)$ we have
\begin{equation}
\int_{u(\epsilon+ i\,0^{+})}^0 du \, \frac{1}{(1-u)} =
\ln\left({\tilde \epsilon}\right),
\end{equation}
so 
\begin{eqnarray}
\label{res2}
\int_{u(\epsilon+ i\,0^{+})}^0 du \, \frac{1}{1-u} + c.c. &=&
\ln\left({\tilde \epsilon}\right) + \ln\left({\tilde \epsilon}\right) \nonumber\\ 
&& = \ln\left( - i\,\sqrt{2 \epsilon} + O(\epsilon) \right) +
\ln\left( i\,\sqrt{2 \epsilon} + O(\epsilon) \right) \nonumber\\
&& = - i\,\frac{\pi}{2} + i\,\frac{\pi}{2} + 2 \ln\left(\sqrt{2 \epsilon}\right)  
+ O(\epsilon)^{1/2} \nonumber\\
&& = \ln 2 + \ln\epsilon + O(\epsilon)^{1/2}
\end{eqnarray}
Note that for constant terms it results
\begin{equation}
\label{res3}
\int_{u(\epsilon+ i\,0^{+})}^0 du \, {\rm const.} + c.c. = - 2 \, {\rm const.}
  + O(\epsilon)^{1/2}.
\end{equation}
Using the results Eq. (\ref{res1}) and Eq. (\ref{res3}), it is easy to show
that the spurious terms of order $(Z \alpha)^2$ in the expression (\ref{resFu})
vanish after the final $u$-integration.

The final result for the high-energy-part ($4P_{1/2}$-state) is
\begin{equation}
F_H(4P_{1/2}) = -\frac{1}{6} + (Z \alpha)^2 
\left[ \frac{24409}{86400} - \frac{499}{720} 
(\ln 2 + \ln \epsilon) - \frac{23}{90\, \epsilon} \right].
\end{equation}
We give here the complete results for the high-energy-parts of the other
states treated in this paper
\begin{equation}
F_H(3P_{1/2}) = -\frac{1}{6} + (Z \alpha)^2 
\left[ \frac{6191}{24300} - \frac{268}{405} 
(\ln 2 + \ln\epsilon) - \frac{20}{81 \epsilon} \right],
\end{equation}
\begin{equation}
F_H(3P_{3/2}) = \frac{1}{12} + (Z \alpha)^2 
\left[ \frac{67903}{194400} - \frac{148}{405} 
(\ln 2 + \ln \epsilon ) - \frac{20}{81 \epsilon} \right],
\end{equation}
\begin{equation}
F_H(4P_{3/2}) = \frac{1}{12} + (Z \alpha)^2 
\left[ \frac{31399}{86400} - \frac{137}{360} 
(\ln 2 + \ln\epsilon) - \frac{23}{90 \epsilon} \right].
\end{equation}
%

\section{The low--energy part}
The low--energy part of the energy shift originates from low--energy
virtual photons. The energy of the photons is comparable
in magnitude to the binding energy of the electron (order $(Z\,\alpha)^2$).
Therefore it is impossible to expand the electron propagator in powers of
the binding field. We have to treat the binding field
non-perturbatively. An expansion in powers of $(Z\,\alpha)$ is
accomplished by considering the spatial momenta of the virtual photon
and the electron momenta as expansion parameters.

Choosing the Coulomb gauge for the photon propagator,
one finds that only the spatial elements of this propagator 
contribute \cite{jp1}. The $\omega$-integration along $C_L$ is performed 
first, which leads to the following expression for $E_L$,
\begin{equation}
\label{defEL}
E_L = - e^2 P \int_{| {\bf k} | < \epsilon}
\frac{d^3 k}{(2 \pi)^3 \, 2 | {\bf k} |} \, \delta^{T, ij} 
\langle \psi | \alpha^i e^{i \, {\bf k} \, \cdot \, {\bf r}}
\frac{1}{H_D - (E_{\psi} - \omega)}  
\alpha^j e^{-i \, {\bf k} \, \cdot \, {\bf r}} | \psi \rangle
\quad (\omega \equiv | {\bf k} |).
\end{equation}
$H_D$ denotes the 
Dirac-Coulomb-Hamiltonian $H_D = \bbox{\alpha}\cdot{\bf p} + \beta\,m+ V$, 
$\delta^T$ is the transverse delta function,
and $\alpha^i$ refers to the Dirac $\alpha$-matrices. The principal
value of the above integral is the real quantity corresponding to the
energy shift in one--loop order. The imaginary part of the $C_L$
integration,
which leads to the decay--width of the state, has been dropped in Eq.
(\ref{defEL}). In the matrix element
\begin{equation}
P^{ij} = 
\langle \psi | \alpha^i e^{i \, {\bf k} \, \cdot \, {\bf r}}
\frac{1}{H_D - (E_{\psi} - \omega)}  
\alpha^j e^{-i \, {\bf k} \, \cdot \, {\bf r}} | \psi \rangle,
\end{equation}
we introduce a unitary Foldy-Wouthuysen transformation $U$,
\begin{equation}
P^{ij} = 
\langle U \psi | 
(U \, \alpha^i e^{i \, {\bf k} \, \cdot \, {\bf r}} \, U^{+})
\frac{1}{U \, (H_D - (E_{\psi} - \omega)) \, U^{+}}  
(U \, \alpha^j e^{-i \, {\bf k} \, \cdot \, {\bf r}} \, 
U^{+}) | U \psi \rangle.
\end{equation}
The lower components of the Foldy-Wouthuysen transformed Dirac wave 
function $\psi$ vanish up to $(Z \alpha)^2$, 
so that we may approximate $| U \psi \rangle$ by
\begin{equation}
| U \psi \rangle = | \phi \rangle + | \delta \phi \rangle \quad
\mbox{with} \quad \langle \phi | \delta \phi \rangle = 0,
\end{equation}
where $| \phi \rangle$ is the non--relativistic 
(Schr\"{o}dinger-Pauli) wave function,
and $| \delta \phi \rangle$ is the relativistic correction.

We define an operator acting on the spinors as even if it does
not mix upper and lower components of spinors, 
and we call the odd operator odd if it mixes upper and lower
components.  The Foldy-Wouthuysen Hamiltonian consists of even 
operators only. For the upper left
$2 \times 2$ submatrix of this Hamiltonian, we find the result 
\cite{itzykson} 
\begin{equation}
H_{\rm FW} = U \, (H_D - (E_{\psi} - \omega)) \, U^{+} = 
m + H_S + \delta H,
\end{equation}
where $H_S$ refers to the Schr\"{o}dinger--Hamiltonian, and $\delta H$ is
is the relativistic correction,
\begin{equation}
\delta H = - \frac{\left({\bf p}\right)^4}{8 \, m^3}
+ \frac{\pi (Z \alpha)}{2 \, m^2} \, \delta({\bf r}) +
\frac{(Z \alpha)}{4 \, m^2 \, r^3} \, {\bbox{\sigma}} \, \cdot \, {\bf L}
\end{equation}
It is interesting to note the reason why we can ignore the lower $2
\times 2$ submatrix of the FW Hamiltonian in our scheme of
calculation. The lower $2 \times 2$ submatrix contains the terms
$-m - (E_{\psi} - \omega) \approx - 2\, m$ as the dominating 
$(Z\, \alpha)^0$ contribution. Therefore the integral vanishes in the
limit $\epsilon \to 0$. This can be checked by considering the
integral in Eq. (\ref{defEL}), inserting a spectral resolution for the 
Dirac Coulomb propagator and performing the integral over $d^3 k$ 
after suitable angular averaging. The upper $2 \times 2$ submatrix has no 
$(Z \, \alpha)^0$ contribution, because terms proportional to 
$m$ and $E_{\psi}$ cancel. This submatrix contributes to the Lamb shift. 
Now we turn to the calculation of the Foldy-Wouthuysen transform
of the operators $\alpha^i \exp\left( {\bf k}\, \cdot \, {\bf r}
\right)$. The expression 
$U \, \alpha^i \exp\left(i {\bf k} \, \cdot \, {\bf r}\right) \, U^{+}$
is to be calculated.  
Assuming that $\omega = | {\bf k} |$ is of the order $O((Z \alpha)^2)$,
we may expand the expression
$U \, \alpha^i \, e^{i {\bf k} \, \cdot \, {\bf r}} \, U^{+}$ 
in powers of $(Z \alpha)$.
The result of the calculation is
\begin{eqnarray}
\label{alphairaw}
U \, \alpha^i e^{i {\bf k} \, \cdot \, {\bf r}} \, U^{+} & = &
\alpha^i \left(1 + i \left( {\bf k} \, \cdot {\bf r} \right) -
\frac{1}{2} \left( {\bf k} \, \cdot \, {\bf r} \right)^2 \right) 
- \frac{1}{2 \, m^2} p^i \, \left( \bbox{\alpha} \, \cdot \, \bbox{p} 
\right) \\
& & + \gamma^0 \, 
 \frac{p^i}{m} \left(1 + i \left( {\bf k} \, \cdot {\bf r} \right) -
\frac{1}{2} \left( {\bf k} \, \cdot \, {\bf r} \right)^2 \right)
\nonumber \\
& & - \gamma^0 \frac{1}{2 \, m^3} p^i {\bf p}^2 -
\frac{1}{2 \, m^2} \, \frac{\alpha}{r^3} \, 
  \left( {\bf r} \times \bbox{\Sigma} \right)^i \nonumber \\
& & + \frac{1}{2 \, m} \, \gamma^0 \, 
  \left( {\bf k} \, \cdot \, {\bf r} \right)
     \left( {\bf k} \times \bbox{\Sigma} \right)^i
- \frac{i}{2 \, m} \, \gamma^0 \,
   \left( {\bf k} \times \bbox{\Sigma} \right)^i \nonumber.
\end{eqnarray}
In the limit $\epsilon \to 0$ the odd
operators in the above expression do not contribute to the self
energy in $(Z \alpha)^2$ relative order, 
because up to $(Z \alpha)^2$ relative order, these operators only join 
the upper components of the wave function with the lower components
of the Dirac Coulomb Hamiltonian. This contribution vanishes, as 
described. So one can neglect the odd operators.
By using symmetry arguments, it can be shown easily that the
last term in the above expression (proportional to 
${\bf k} \times \bbox{\Sigma}$) 
does not contribute to the Lamb shift in $(Z \alpha)^2$
relative order for $\epsilon \to 0$, either. 

Because we can ignore odd operators, and because
the lower components of the Foldy-Wouthuysen transformed wave function
vanish, we keep only the upper left $2 \times 2$ submatrix of Eq. 
((\ref{alphairaw})), and we write 
$U \, \alpha^i \, e^{i {\bf k} \, \cdot \, {\bf r}} U^{+}$ as
\begin{eqnarray}
\label{alphaitransformed}
U \, \alpha^i e^{i {\bf k} \, \cdot \, {\bf r}} \, U^{+} & \simeq &
\frac{p^i}{m} \left(1 + i \left( {\bf k} \, \cdot {\bf r} \right) -
\frac{1}{2} \left( {\bf k} \, \cdot \, {\bf r} \right)^2 \right) \\
& & - \frac{1}{2 \, m^3} p^i {\bf p}^2 -
\frac{1}{2 \, m^2} \, \frac{\alpha}{r^3} \, 
  \left( {\bf r} \times \bbox{\sigma} \right)^i \nonumber \\
& & + \frac{1}{2 \, m} \left( {\bf k} \, \cdot \, {\bf r} \right)
 \left( {\bf k} \times \bbox{\sigma} \right)^i\,, \nonumber
\end{eqnarray}
This can be rewritten as
\begin{equation}
U \, \alpha^i \, e^{i {\bf k} \, \cdot \, {\bf r}} \, U^{+} =
\frac{p^i}{m} \, e^{i {\bf k} \, \cdot \, {\bf r}} + \delta y^i,
\end{equation}
where $\delta y^i$ is of order $(Z \alpha)^3$. It is understood
that the term $\frac{p^i}{m} \, e^{i {\bf k} \, \cdot \, {\bf r}}$ is 
also expanded up to the order $(Z \alpha)^3$. 
Denoting the Schr\"{o}dinger energy by $E$
($E = - (Z \alpha)^2 \, m / n^2$ for nP states) and the
first relativistic correction to $E$ by $\delta E$, we can thus 
write the matrix element $P^{ij}$ as
\begin{equation}
\label{pij}
P^{ij} = 
\langle \phi + \delta \phi | \left[ \frac{p^i}{m} \, 
  e^{i {\bf k} \, \cdot \, {\bf r}} + \delta y^i \right] \,
\frac{1}{H_S - (E - \omega) + \delta H - \delta E} \,
\left[ \frac{p^j}{m} \, 
  e^{- i {\bf k} \, \cdot \, {\bf r}} + \delta y^j \right] 
| \phi + \delta \phi \rangle.
\end{equation}
We now define the dimensionless quantity
\begin{equation}
\label{definitionofP}
P = \frac{m}{2} \, \delta^{T, ij} \, P^{ij}.
\end{equation}
Up to $(Z \alpha)^2$, we can write the matrix element $P$ 
as the sum of the contributions
((\ref{pnd}), (\ref{pnq}), (\ref{pdeltay}), (\ref{pdeltah}), (\ref{pdeltae}), 
(\ref{pdeltaphi})). 
The leading contribution (the  ``non-relativistic dipole'') is given by
\begin{equation}
\label{pnd}
P_{\rm nd} = \frac{1}{3 m} \, \langle \phi | p^i \, 
\frac{1}{H_S - (E - \omega)} \, p^i | \phi \rangle.
\end{equation}
The other contributions to $P$ are \cite{jp1}
\begin{itemize}
\item the non-relativistic quadrupole,
\begin{equation}
\label{pnq}
P_{\rm nq} = \frac{1}{3 m} \, \langle \phi | p^i \, 
e^{ i {\bf k} \, \cdot \, {\bf r}} \, 
\frac{1}{H_S - (E - \omega)} \, p^i \, 
e^{ - i {\bf k} \, \cdot \, {\bf r}} | \phi \rangle - P_{\rm nd},
\end{equation}
\item the corrections to the current $\alpha^i$ from the
Foldy-Wouthuysen transformation,
\begin{equation}
\label{pdeltay}
P_{\delta y} = \delta^{T, ij} \, 
\langle \phi | \delta y^i \, 
\frac{1}{H_S - (E - \omega)} \, p^j \, 
e^{ - i {\bf k} \, \cdot \, {\bf r}} | \phi \rangle,
\end{equation}
\item the contribution due to the relativistic Hamiltonian,
\begin{equation}
\label{pdeltah}
P_{\delta H} = - \frac{1}{3 m} \, \langle \phi | p^i \, 
\frac{1}{H_S - (E - \omega)} \, \delta H \, 
\frac{1}{H_S - (E - \omega)} \, p^i | \phi \rangle,
\end{equation}
\item the contribution due to the relativistic correction to the
energy,
\begin{equation}
\label{pdeltae}
P_{\delta E} = \frac{1}{3 m} \, \langle \phi | p^i \, 
\frac{1}{H_S - (E - \omega)} \, \delta E \, 
\frac{1}{H_S - (E - \omega)} \, p^i | \phi \rangle,
\end{equation}
\item and due to the relativistic correction
to the wave function,
\begin{equation}
\label{pdeltaphi}
P_{\delta \phi} = \frac{2}{3 m} \, \langle \delta \phi | p^i \, 
\frac{1}{H_S - (E - \omega)} \, p^i | \phi \rangle.
\end{equation}
\end{itemize}
Almost all of the above contributions are calculated using a 
coordinate space representation of the Schr\"{o}dinger Coulomb
propagator given in \cite{jp1} and \cite{swainsondrake}. Formulae
given in \cite{bateman} and \cite{buchholz} prove useful for
the summation over the intermediate quantum numbers. For the 
non-relativistic quadrupole contribution, however, we use the 
momentum space representation due to Schwinger \cite{lieber}
\begin{eqnarray}
G({\bf p},{\bf p}',\Omega) &=& 4 \pi m X^3 \left( \frac{i 
e^{i\pi\tau}}{2\sin \pi\tau}\right)\int_{1}^{0^+}d\rho\,
\rho^{-\tau}\, \frac{d}{d\rho} \nonumber\\
&& \times \frac{1-\rho^2}{\rho}\frac{1}{\Bigl[X^2 ({\bf p}-{\bf p}')^2 +
({\bf p}^2 + X^2)({\bf p}'^2 + X^2) \,(1-\rho)^2/(4\rho)\Bigr]^2},
\end{eqnarray}
where $X=\sqrt{-2m\Omega}$, $\tau = m\alpha /X$. For non-integer
$\tau$, one may replace the complex integration around the origin
(in the positive sense) by a much simpler integral
\begin{equation}
\left( \frac{i e^{i\pi\tau}}{2\sin \pi\tau}\right)
\int_{1}^{0^+}d\rho \, \rho^{-\tau} \, g(\rho) \to 
\int_0^1 d\rho \, \rho^{-\tau} \, g(\rho).
\end{equation}
We then perform the calculation of $P_{\rm NQ}$ in momentum 
space using formulae given in the paper by
Gavrila and Costescu \cite{gavrilacostescu1}.
The momentum space wave functions of P states are given in 
\cite{lieber}.
Calculations for $P_{\rm NQ}$ become increasingly complex. It
should be noted that in the case of the 4P wave 
function, one has to deal
with intermediate expressions of up to 20,000 
terms. We do not describe these calculations in any further detail.
Evaluations for this
part of the calculation were done on {\sc IBM RISC/6000} systems with
the help of the computer algebra system {\sc Mathematica} 
\cite{mathematica}.

The final $\omega$ integration is done by change of variable
\begin{equation}
\label{cov}
\omega \to t, \quad \mbox{where} \quad t = \frac{1}{\sqrt{1 + 
(2 \, n^2 \, \omega)/((Z\, \alpha)^2\, m)}},
\end{equation}
($t=0$ corresponds to $\omega = \infty$, and $t=1$ corresponds to 
$\omega=0$). As an example for a $P$ matrix element, we give here the
result for a contribution to $P_{\delta H}({\rm 4P}_{1/2})$, caused by
the Russell-Saunders coupling term in the Foldy-Wouthuysen transformed
Dirac Hamiltonian:
\begin{eqnarray}
\label{PLS}
P_{{\bf L}\cdot{\bf S}}(4{\rm P}_{1/2}) &=&
\langle \phi | p^i \, \frac{1}{H_S - (E_{\phi} - \omega)} \,
\bigg[ \frac{\alpha}{4\, m^2 r^3} \, \bbox{\sigma}\cdot {\bf L} \bigg] \,
\frac{1}{H_S - (E_{\phi} - \omega)} \, p^i | \phi \rangle \nonumber\\ 
&=& (Z \alpha)^2 \,
\bigg[16384\,F_{366}(t)\,t^6\,(-1 + 2\,t)\,(1 + 2\,t)\,
(-1 + 4\,t)\,(1 + 4\,t)\,(-5 + 9\,t^2)^2\bigg]\bigg/ \nonumber\\
&& \hspace{3ex} \bigg[(675\,(-1 + t)^2\,(1 + t)^{14}\bigg] \nonumber\\
&& + (Z \alpha)^2 \, \bigg[8192\,\Phi_{23}(t)\,t^6\,(-1 + 2\,t)\,
(1 + 2\,t)\,(-1 + 4\,t)\,(1 + 4\,t)\,(-5 + 9\,t^2)^2\bigg]\bigg/
\nonumber\\
&& \hspace{3ex} \bigg[675\,(-1 + t)^2\,(1 + t)^{14}\bigg] \nonumber\\
&& - (Z \alpha)^2 \, \bigg[16384\,\Psi_{63}(t)\,t^6\,(-1 + 2\,t)\,
(1 + 2\,t)\,(-1 + 4\,t)\,(1 + 4\,t)\,\nonumber\\
&& \hspace{3ex} \times (-5 + 9\,t^2)^2\bigg]\bigg/
\bigg[675\,(-1 + t)^2\,(1 + t)^{14}\bigg]
\nonumber\\
&& - (Z \alpha)^2 \, \bigg[16\,F_4(t)\,t^5\,(-5 + 9\,t^2)\,(5 +
9\,t^2)\bigg] \bigg/
\bigg[45\,(-1 + t)^5\,(1 + t)^5\bigg] \nonumber\\
&& 
- (Z \alpha)^2 \bigg[4096\,\gamma\,t^7\,(1 + 2\,t)\,(1 + 4\,t)\,
     (-5 + 9\,t^2)^2 \times \nonumber\\
&& \times \bigg(-630 + 2988\,t - 22677\,t^2 + 322539\,t^3 - 
       1308681\,t^4 + 5929071\,t^5 - \nonumber\\
&& \hspace{3ex} + 20884413\,t^6 + 53752275\,t^7 - 
       124047785\,t^8 + 233683811\,t^9 \nonumber\\
&& \hspace{3ex} - 362131987\,t^{10} + 489361437\,t^{11} - 
       552966475\,t^{12} + 526774701\,t^{13} \nonumber\\
&& \hspace{3ex} - 433556447\,t^{14} + 298335889\,t^{15} - 
       171601869\,t^{16} + 84114837\,t^{17} \nonumber\\
&& \hspace{3ex} - 33425148\,t^{18} + 10588500\,t^{19} - 
       2811264\,t^{20} + 536448\,t^{21} \nonumber\\
&& \hspace{3ex} - 64512\,t^{22} + 9216\,t^{23}\bigg) \bigg] \bigg/ \nonumber\\
&& \hspace{3ex} \bigg[ 675\,(-2 + t)\,(-1 + t)^8\,(1 + t)^24\,(-3 + 2\,t)\,(-7 + 4\,t)\,(-5 + 4\,t)\,
     (-3 + 4\,t)\bigg] \nonumber\\
&& + (Z \alpha)^2 \, \bigg[t^4\,\bigg(54385222500 + 13984771500\,t + 
       15573367761525\,t^2 \nonumber\\
&& \hspace{3ex} + 137704964619600\,t^3 - 1186676073750825\,t^4 - 
       891514989328950\,t^5 \nonumber\\
&& \hspace{3ex} - 18753057125404875\,t^6 + 
       257731557911828220\,t^7 - 644138132939443685\,t^8 \nonumber\\
&& \hspace{3ex} - 162189845176905698\,t^9 + 5703957333855251853\,t^{10} \nonumber\\
&& \hspace{3ex} - 40461889136476779376\,t^{11} + 
179879463449806219483\,t^{12}  \nonumber\\
&& \hspace{3ex} - 
       470465622621377811686\,t^{13} + 722355119079050313441\,t^{14} \nonumber\\
&& \hspace{3ex}  - 285989274333023499852\,t^{15} -
2287627466701513790137\,t^{16} \nonumber\\
&& \hspace{3ex} + 
       8721622433626707698710\,t^{17} -
17878002921800012898021\,t^{18} + \nonumber\\
&& \hspace{3ex} +22584212519300184263992\,t^{19} -
11732105237171234433515\,t^{20} \nonumber\\
&& \hspace{3ex} - 22234628025915862014322\,t^{21} + 73751337137966874205423\,t^{22} \nonumber\\
&& \hspace{3ex} -118367926111467660583708\,t^{23} +
122509973591672089703233\,t^{24}  
\nonumber\\
&& \hspace{3ex} - 
       68708753123836196116438\,t^{25}  - 27553239135523122581409\,t^{26} \nonumber\\
&& \hspace{3ex}
+ 126820257213436783378720\,t^{27}  - 189954960243232498928039\,t^{28}\nonumber\\
&& \hspace{3ex} + 
       199312481720191257908238\,t^{29}  - 164901212087123003552709\,t^{30} \nonumber\\
&& \hspace{3ex}+ 
       112286946900261699896044\,t^{31}  - 64112830789916243409479\,t^{32} \nonumber\\
&& \hspace{3ex}+
30834230557497197343734\,t^{33} 
- 12455975426801658077884\,t^{34} \nonumber\\
&& \hspace{3ex} + 
       4211137005166537183048\,t^{35}  - 1183248309588246704096\,t^{36}\nonumber\\
&& \hspace{3ex} + 
       270125031800068986496\,t^{37} -
48193573673016712704\,t^{38} \nonumber\\
&& \hspace{3ex} + 6583393931443034112\,t^{39} -
691086520295792640\,t^{40}  \nonumber\\
&& \hspace{3ex} + 41007381492105216\,t^{41} +
1314475331420160\,t^{42}\bigg)\bigg]\bigg/ \nonumber\\
&& \bigg[425250\,(-2 + t)^2\,(-1 + t)^{10}\,(1 + t)^{24}\,(-3 + 2\,t)^2 \nonumber\\
&& \hspace{3ex} \times (-1 + 2\,t)\,
     (-7 + 4\,t)^2\,(-5 + 4\,t)^2\,(-3 + 4\,t)^2\,(-1 + 4\,t)\bigg] \nonumber\\ 
&& + (Z \alpha)^2  \, 
\bigg[8192\,t^8\,(-1 + 2\,t)\,(-1 + 4\,t)\,(-5 + 9\,t^2)^2\,(3 + 6\,t +
3\,t^2 + 8\,t^3) \nonumber\\ 
&& \hspace{3ex} \times \ln(2/(1 + t))\bigg]\bigg/
\bigg[675\,(-1 + t)^{12}\,(1 + t)^6\bigg] \nonumber\\ 
&& + (Z \alpha)^2  \bigg[16\,G_4(t)\,t^5\,(-5 + 9\,t^2)\,
     (-1125 + 2400\,t + 145390\,t^2 \nonumber\\
&& \hspace{3ex} - 19200\,\gamma\,t^2 - 148320\,t^3 - 
       2026412\,t^4 + 418560\,\gamma\,t^4 \nonumber\\
&& \hspace{3ex} + 1027200\,t^5 + 7316050\,t^6 - 
       1920000\,\gamma\,t^6 - 1382400\,t^7 \nonumber\\
&& \hspace{3ex} - 7444143\,t^8 + 
       2211840\,\gamma\,t^8 - 19200\,t^2\,\ln(2/(1+t)) + 
       418560\,t^4\,\ln(2/(1+t)) \nonumber\\
&& \hspace{3ex} - 1920000\,t^6\,\ln(2/(1+t)) + 
       2211840\,t^8\,\ln(2/(1+t)))\bigg]\bigg/\nonumber\\
&& \hspace{3ex} \bigg[10125\,(-1 + t)^8\,(1 + t)^8 \bigg] \nonumber\\
&& + (Z \alpha)^2 \, \bigg[256\,t^7\,(-5 + 9\,t^2)^2\,
   (8 + 45\,t - 305\,t^2 - 175\,t^3 + 283\,t^4 \nonumber\\
&& \hspace{3ex}- 
       265\,t^5 + 3421\,t^6 + 395\,t^7 + 433\,t^8)\,\ln[(2\,t)/(1 +
t)]\bigg]\bigg/ \nonumber\\
&& \hspace{3ex} \bigg[675\,(-1 + t)^12\,(1 + t)^7\bigg],
\end{eqnarray}
where
\begin{equation}
F_4(t) = {_2}F_1\left[1, -4t,1-4t,((t-1)/(t+1))^2\right],
\end{equation}
\begin{equation}
G_4(t) = {_2}F_1\left[1, -4t,1-4t,(t-1)/(t+1)\right],
\end{equation}
\begin{equation}
F_{366}(t) = t^2 \, \sum^{\infty}_{k=6} \,
\frac{\left((t-1)/(t+1)\right)^k}{3 - 4 t + k} \,
\frac{\partial}{\partial b}({_2}F_1)[-k,6,62/(1+t)],
\end{equation}
\begin{equation}
\Phi_{23}(t) = \sum_{k=6}^{\infty} \,
\frac{\left((t-1)/(t+1)\right)^{2 k}}{(3 - 4 t + k)^2},
\end{equation}
\begin{equation}
\Psi_{63}(t) = \sum_{k=6}^{\infty} \,
\frac{\left((t-1)/(t+1)\right)^{2 k}}{(3 - 4 t + k)} \, \Psi(k + 6),
\end{equation}
where $\Psi$ denotes the logarithmic derivative of the $\Gamma$
function.
The quadratic singularity in the result for 
$P_{{\bf L} \cdot {\bf S}}(4{\rm P}_{1/2})$ in Eq. (\ref{PLS}) at $t=3/4$ 
(given by the $(-3 + 4t)^2$--term in the denominator of the purely
rational function) corresponds to the decay into the 3D
state. One can check explicitly that the insertion of corresponding
intermediate states in the spectral decomposition of the propagators 
necessitates the existence of quadratic singularities in the $P$
matrix elements, and that
the quadratic singularities occur only in the matrix elements with two
propagators. The quadratic singularities are a
consequence of the perturbative treatment of $\delta H$ in the
propagator $1/(H_S + \delta H - (E_{\phi} + \delta E - \omega))$
(expansion in $\delta H$ is not allowed in the vicinity of a
pole of the resolvent $G(E) = 1/(H_S - E)$). The
integration procedure for the quadratic singularities is as follows: 
first we isolate and calculate analytically the integral of 
the term that gives
rise to the singularity (as a function of $t$), then we take the
difference
of the edge terms at $t=1$ and $t=0$. This procedure takes back the
effect of the perturbative treatment and assigns the correct
value to the $t$ integral. As the final step, we subtract the term
that
gave rise to the quadratic singularity and proceed with the rest of
the terms in the usual way described in \cite{jp1}. 

The integration procedure deserves some further comments. We also
encounter
in the matrix elements singularities of linear type at $t=1/4$,
$t=1/2$ and $t=3/4$, which also correspond to the decay of the excited
state. 
The residue taken at these singularities yields the decay width of the
respective states. In order to obtain the principal value of the
$t$-integral, one has to symmetrize the integrand around all the 
singularities. This is also accomplished by symbolic procedures
written in the computer algebra language {\sc Mathematica}.

The results of the calculations have been checked in many ways.
An important cross-check is the cancellation of $\epsilon$-divergent
terms in the sum of the high and low--energy parts. By considering the
expansion of the propagators in powers of $1/\omega$, the logarithmic
singularities of all contributions can be calculated individually and
agree with the results obtained from complete evaluation.
 
The contributions to the low--energy part $F_L$ in 
$(Z \alpha)^2$ relative order are given in Table \ref{table3P} and
Table \ref{table4P}.
Summing all contributions, 
we obtain the following complete results for the contribution of the low
energy parts $F_L$:
\begin{equation}
F_L(3P_{1/2}) = - \frac{4}{3} \ln k_0(3{\rm P}) +
(Z \alpha)^2 \, \bigg[-0.94378(1) + \frac{20}{81 \, \epsilon} 
+ \frac{268}{405} \, \ln
\frac{\epsilon}{(Z \alpha)^2}\bigg],
\end{equation}
\begin{equation}
F_L(3P_{3/2}) = - \frac{4}{3} \ln k_0(3{\rm P}) +
(Z \alpha)^2 \, \bigg[-0.69356(1) + \frac{20}{81 \, \epsilon} 
+ \frac{148}{405} \, \ln
\frac{\epsilon}{(Z \alpha)^2}\bigg],
\end{equation}
\begin{equation}
F_L(4P_{1/2}) = - \frac{4}{3} \ln k_0(4{\rm P}) +
(Z \alpha)^2 \, \bigg[-0.99780(1) + \frac{23}{90 \, \epsilon} \, 
+ \frac{499}{720} \, \ln
\frac{\epsilon}{(Z \alpha)^2}\bigg],
\end{equation}
and
\begin{equation}
F_L(4P_{3/2}) = - \frac{4}{3} \ln k_0(4{\rm P}) +
(Z \alpha)^2 \, \bigg[-0.73057(1) + \frac{23}{90 \, \epsilon} 
+ \frac{137}{360} \, \ln
\frac{\epsilon}{(Z \alpha)^2}\bigg].
\end{equation}

\section{RESULTS AND EVALUATION OF THE LAMB SHIFT}

Summing the contributions from the high and low--energy parts,
we obtain the following results for the scaled $F$ function
defined in Eq. (\ref{defF}): 
\begin{equation}
F(3{\rm P}_{1/2}) = -\frac{1}{6} - \frac{4}{3} \ln k_0(3{\rm P}) + (Z \alpha)^2
\, \Biggl[-1.14768(1) + \frac{268}{405} \ln[(Z \alpha)^{-2}] \Biggr],
\end{equation}
\begin{equation}
F(3{\rm P}_{3/2}) = \frac{1}{12} - \frac{4}{3} \ln k_0(3{\rm P}) + (Z \alpha)^2
\, \Biggl[-0.59756(1) + \frac{148}{405} \ln[(Z \alpha)^{-2}] \Biggr],
\end{equation}
\begin{equation}
F(4{\rm P}_{1/2}) = -\frac{1}{6} - \frac{4}{3} \ln k_0(3{\rm P}) + (Z \alpha)^2
\, \Biggl[-1.19568(1) + \frac{499}{720} \ln[(Z \alpha)^{-2}] \Biggr],
\end{equation}
\begin{equation}
F(4{\rm P}_{3/2}) = \frac{1}{12} - \frac{4}{3} \ln k_0(3{\rm P}) + (Z \alpha)^2
\, \Biggl[-0.63094(1) + \frac{137}{360} \ln[(Z \alpha)^{-2}] \Biggr].
\end{equation}
The results obtained for $A_{4,0}$ and $A_{6,1}$ are in agreement with those
previously known \cite{sapirsteinyennie1}. The values of the Bethe
logarithms \cite{km1}, \cite{ds1}
\begin{equation}
\ln k_0 (3{\rm P}) = -0.03819023(1), \quad 
\ln k_0 (4{\rm P}) = -0.04195489(1)
\end{equation}
could be verified numerically with a 7 figure accuracy from our
analytic expressions by numerical (Gaussian) integration.
For $A_{6,1}$, we use the following
general formula which may be extracted from the work by Erickson and
Yennie (\cite{ericksonyennie12}, Eq. (4.4a) ibid., upon subtraction of
the vacuum polarization contribution implicitly contained in the
quoted equation):
\begin{eqnarray}
A_{6,1}(n,l,j) &=& \frac{4}{3} \Biggl[  \left(1 -
\delta_{l,0}\right) \,
\frac{8\, 
\left(3 - (l(l+1))/n^2\right)}{(2\,l - 1) \, (2\,l) \, (2\,l+1) \, 
(2\,l + 2)\,(2\,l+3)} \nonumber\\
&& + \delta_{l,1} \, 
\biggl[1-\frac{1}{n^2}\biggr] \, \biggl[\frac{1}{10} + \frac{1}{4} \,
\delta_{j,l-1/2} \biggr] \nonumber\\
&& + \delta_{l,0} \, \biggl[-\frac{601}{240} - \frac{77}{60 \, n^2} + 
7 \ln 2 + 3 \left( \gamma - \ln n + \Psi(n+1) \right) \biggr] \Biggr],
\end{eqnarray}
where $\gamma$ is Euler's constant, and $\Psi$ refers to the
logarithmic derivative of the $\Gamma$ function.
For P states ($l=1$), this formula reduces to
\begin{equation}
A_{6,1}(n,1,j) = \frac{4}{3} \left[\frac{1}{15} \left(3-\frac{2}{n^2}\right) +
\left(1-\frac{1}{n^2} \right)\,\left(\frac{1}{10} + \frac{1}{4}\,
\delta_{j,1/2}\right)\right]
\end{equation}
and is in agreement with our results. From our results for $F$, we
extract the following new values for the coefficients $A_{6,0}$:
\begin{equation}
A_{6,0}(3{\rm P}_{1/2}) = -1.14768(1), \quad 
A_{6,0}(3{\rm P}_{3/2}) = -0.59756(1),
\end{equation}
and
\begin{equation}
A_{6,0}(4{\rm P}_{1/2}) = -1.19568(1), \quad 
A_{6,0}(4{\rm P}_{3/2}) = -0.63094(1).
\end{equation}
The new results for $A_{6,0}$ are the main results of this work.
They are in excellent
agreement with data obtained from numerical calculations by one of the
authors (PJM) and Y. K. Kim \cite{mohrkim}. Mohr and Kim
calculate the $F$ function defined in (\ref{defF}) numerically for
$Z \geq 10$, treating the binding field non-perturbatively. By
extrapolating their numerical data \cite{mohrkim} to the region
of small $Z$, we obtain the following estimates for the remainder
function $G_{{\rm SE},7}$ implicitly defined in Eq. (\ref{defF}).
\begin{equation}
G_{{\rm SE},7}(3{\rm P}_{1/2}, Z=1) = 3.6 \pm 0.5, \quad 
G_{{\rm SE},7}(3{\rm P}_{3/2}, Z=1) = 2.6 \pm 0.5,
\end{equation}
and
\begin{equation}
G_{{\rm SE},7}(4{\rm P}_{1/2}, Z=1) = 3.9 \pm 0.5, \quad 
G_{{\rm SE},7}(4{\rm P}_{3/2}, Z=1) = 2.8 \pm 0.5.
\end{equation}
The uncertainties in $G_{{\rm SE},7}$ are used to estimate the
theoretical uncertainty from the one--loop contribution. When modeling
the numerical data, it must be taken into account that as noted
by Karshenboim \cite{karshenboim2}, $A_{7,1}$ coefficients
vanish for P states. Values of
$A_{6,0}$ and $G_{{\rm SE},7}$ for 2P states are given in \cite{jp1}.  
We use the following implicit definition of the Lamb shift ${\cal L}$:
\begin{equation}
\label{defElamb}
E = m_r \left[ f(n,j)-1 \right] - \frac{m_r^2}{2 (m + m_N)}
\left[ f(n,j) - 1 \right]^2 + {\cal L} +
E_{\rm hfs},
\end{equation}
where $E$ is the energy level of the two-body-system 
and $f(n,j)$ is the dimensionless Dirac energy, $m$
is the electron mass, $m_r$ is the reduced mass of the system 
and $m_N$ is the nuclear mass. It should be noted that we 
consider the hfs--fs mixing term as a contribution to the
hyperfine structure. The small hfs-fs-mixing correction which is discussed 
in \cite{pipkin1}, mixes the $F=1$ sublevels of the ${\rm P}_{1/2}$ and
${\rm P}_{3/2}$ states and shifts the center of the hyperfine
levels. It should be taken into account when the 
fine structure is deduced from precision experiments.

In order to calculate the Lamb shift, we include the
Barker-Glover correction to hydrogen energy levels \cite{barkerglover1},
which we refer to as the $(Z \alpha)^4$ recoil correction. 
We also include the $(Z \alpha)^5$ recoil correction 
calculated by E. Salpeter \cite{salpeter1},
and the results \cite{golosov} for recoil
corrections of order $(Z \alpha)^6 \, m_r/m_N$. 
The results for recoil corrections of order $(Z \alpha)^6 \, m_r/m_N$
have been confirmed by K. Pachucki (see ch. 5 of \cite{jp1}).
We include contributions from 
the higher order two--loop correction of order $(\alpha/\pi)^2 \,
(Z \alpha)^6 \, \ln^2(Z \alpha)^{-2}$ corresponding to the
$B_{6,2}$--coefficient \cite{karshenboim1}, and for
three--loop corrections in lowest order. The theoretical error 
from the two--loop contribution ($B_{6,1}$ and higher terms) 
is estimated as half the contribution from the recently calculated
$B_{6,2}$--coefficient. The higher order
contributions due to vacuum polarization of order $\alpha/\pi 
\, (Z \alpha)^6$ can be obtained by analyzing the small distance 
behavior of the Dirac wave function, i.e. by evaluating the matrix
element of the Uehling potential (see e.g. \cite{manakov1}),
\begin{equation}
V_{\rm VP}({\bf r}) = \frac{\alpha \, (Z \alpha)}{m^2} \, \left[ 
  - \frac{4}{15} \delta({\bf r}) - \frac{1}{35} \, 
    \frac{\bbox{\nabla}^2}{m^2} \delta({\bf r}) + 
       O\left(\bbox{\nabla}^4 \delta({\bf r})\right) \right] \ ,
\end{equation}
with P-state wavefunctions expanded in powers of $Z\alpha$. 
We obtain the results  
\begin{equation}
A^{\rm vac}_{60}(n{\rm P}_{1/2}) = -\frac{3}{35} \,
\frac{n^2-1}{n^2}
\end{equation}
and
\begin{equation}
A^{\rm vac}_{60}(n{\rm P}_{3/2}) = -\frac{2}{105} 
\frac{n^2-1}{n^2} 
\end{equation}
for the leading term.  We have also evaluated the contribution of the
Uehling potential numerically without expansion in $Z\alpha$ with the 
results
\begin{equation}
G_{{\rm U},7}(3{\rm P}_{1/2}, Z=1) = 0.0455, \quad 
G_{{\rm U},7}(3{\rm P}_{3/2}, Z=1) = 0.0249,
\end{equation}
and
\begin{equation}
G_{{\rm U},7}(4{\rm P}_{1/2}, Z=1) = 0.0480, \quad 
G_{{\rm U},7}(4{\rm P}_{3/2}, Z=1) = 0.0262
\end{equation}
where the function $G_{{\rm U},7}$ is defined in analogy with 
$G_{{\rm SE},7}$.  
The contribution of the higher-order terms is negligible compared
to the uncertainty in the higher-order self energy terms.
The Wichmann-Kroll vacuum polarization contribution
is expected to be of order $(Z\alpha)^2$ times the Uehling correction
and is not included here.

The above mentioned contributions to the Lamb shift are listed in
the Tables \ref{lamb3p} and \ref{lamb4p} for the states under 
investigation. It should be noted that the reduced mass dependence
of the terms must be restored in low $Z$ systems to obtain the 
correct value for the Lamb shift. Terms which are caused by the
anomalous magnetic moment of the electron acquire a factor
$(m_r/m_e)^2$ (where $m_r$ is the reduced mass of the system and 
$m_e$ the mass of the electron), all other contributions to the Lamb shift
acquire a factor $(m_r/m_e)^3$. In addition, the argument of the logarithms
$\ln[(Z\,\alpha)^{-2}]$ must be replaced by 
$\ln[(m_e/m_r \, Z\,\alpha)^{-2}]$.
The relevant formulae are also given in \cite{sapirsteinyennie1}).

It should also be noted that for two--loop and three--loop corrections in 
respective lowest order $(\alpha/\pi)^2 \, (Z \alpha)^4$ and 
$(\alpha/\pi)^3 \, (Z \alpha)^4$, only the anomalous magnetic moment of
the electron contributes to the Lamb shift for P states, because the
Dirac form factor $F_1(q^2)$ is infrared finite 
in two-- and three--loop order. So it is only the contribution
from the magnetic form factor $F_2(q^2 = 0)$ which persists. 
The calculation of the contribution
to the Lamb shift is then straightforward.

We obtain the following theoretical results for the Lamb shift of
3P and 4P states:
\begin{equation}
{\cal L}(3{\rm P}_{1/2}) = -3473.75(3) \, {\rm kHz},
\end{equation}
\begin{equation}
{\cal L}(3{\rm P}_{3/2}) = 4037.75(3)  \,{\rm kHz},
\end{equation}
\begin{equation}
{\cal L}(4{\rm P}_{1/2}) = -1401.52(1)  \,{\rm kHz},
\end{equation}
\begin{equation}
{\cal L}(4{\rm P}_{3/2}) = 1767.30(1)  \,{\rm kHz}.
\end{equation}
The theoretical values for the fine structure splitting, using the
1987 Cohen--Taylor value of $\alpha^{-1} = 137.0359895(61)$ 
\cite{cohentaylor1}, are as follows: 
\begin{equation}
{\Delta E}_{\rm fs}(3 {\rm P}) = 3250089.8(3)  \,{\rm kHz}  
\end{equation}
\begin{equation}
{\Delta E}_{\rm fs}(4 {\rm P}) = 1371130.0(1)  \,{\rm kHz}  
\end{equation}
For $2 {\rm P}$ states, the theoretical value is  
${\Delta E}_{\rm fs}(2 {\rm P}) = 10969043(1) \, {\rm kHz}$ \cite{jp1}.
The uncertainty in the theoretical
values for the fine structure splitting is given by the uncertainty in
$\alpha$. Any determination of the fine structure beyond the quoted
uncertainty would yield a value of $\alpha$ improved with respect to
the 1987 value. Given the scattering of available data for $\alpha$ 
\cite{kinoshita2}, such a determination could be useful for checking
the consistency of measurements coming from different fields of physics.

The formula for the fine structure as 
a function of $\alpha$ for atomic hydrogen ($Z = 1$) is as follows: 
\begin{eqnarray}
\Delta E_{\rm fs}(n) &=& E(n{\rm P}_{3/2}) - E(n{\rm P}_{1/2}) \nonumber\\ 
&=& R_{\infty} \bigg[
 \alpha^2 \bigg( \frac{y}{2\,n^3} - \frac{x^2\,y^3}{2\,n^3} \bigg) +
 \alpha^3 \bigg( \frac{y^2}{2 \, \pi \, n^3} \bigg) \nonumber\\
&& \hspace{3ex} + \alpha^4 \Bigg(\frac{a^{(2)}_e \, y^2}{\pi^2 \, n^3} +
  \frac{7\,y}{32\,n^3} + \frac{9\,y}{16\,n^4} - \frac{3\,y}{4\,n^5} +
  \frac{x\,y^2}{4\,n^5\,(1+x)} \Bigg) \nonumber\\
&& \hspace{3ex} + \alpha^5 \Bigg( 
\frac{a^{(3)}_e \, y^2}{\pi^3 \, n^3} + \frac{2\,y^3}{\pi\, n^3}
\left(\Delta A_{6,0}(n) + \frac{1}{15} \,\frac{n^2-1}{n^2} - 
\frac{1}{3}\,\frac{n^2-1}{n^2}\,\ln(y^{-1}\,\alpha^{-2})\right) 
\Bigg) \nonumber\\
&& \hspace{3ex} + \alpha^6 \Bigg( 
\frac{2\,y^3}{\pi\,n^3} \, \Delta G(n) + 
\frac{31 \, y}{256\,n^3} + \frac{45 \, y}{128\,n^4} +
\frac{7 \, y}{64\,n^3} - \frac{45 \, y}{32\,n^6} +
\frac{15 \, y}{16\,n^7} \Bigg) \Bigg] 
+ \delta {\cal L}_{\rm th}(n)
\end{eqnarray}
where the theoretical uncertainty in the difference of the 
Lamb shift of nP states is given by
\begin{equation}
\delta {\cal L}_{\rm th}(2) = 80 \, {\rm Hz},  \quad
\delta {\cal L}_{\rm th}(3) = 30 \, {\rm Hz},  \quad
\delta {\cal L}_{\rm th}(4) = 10 \, {\rm Hz}.
\end{equation}
The mass ratios are
\begin{equation}
x = m_e/m_p \quad \mbox{and} \quad y = m_r/m_e = 1/(1+x).
\end{equation}
$\Delta A_{6,0}(n)$ and $\Delta G(n)$ are defined as
\begin{equation}
\Delta A_{6,0}(n) = A_{6,0}(n{\rm P}_{3/2}) -  A_{6,0}(n{\rm P}_{1/2}),
\quad \quad \Delta G(n) =  G_{\rm SE,7}(n{\rm
P}_{3/2})-G_{\rm SE,7}(n{\rm P}_{1/2}).
\end{equation}
For practical purposes, the $n$-dependence of $\Delta G_{\rm SE,7}(n)$
may be suppressed, because it is a very small contribution (in the 1 Hz
range), and we may assume  $\Delta G_{\rm SE,7}(n) \approx -1.0$.
The two-- and three--loop coefficients to the anomalous magnetic
moment are given by \cite{kinoshita2}
\begin{equation}
a^{(e)}_2 = -0.328478965 \quad \mbox{and} \quad a^{(e)}_3 = 1.18124156.
\end{equation}

\section{conclusions}

The analytic calculation of higher order binding corrections to
the Lamb shift of excited P states has been described in this paper.
We provide more accurate theoretical values of the Lamb shift for 
3P and 4P states in hydrogenlike systems. 

We also give a formula for the fine structure as a
function of $\alpha$ (for 2P, 3P and 4P states), which may be used to 
determine $\alpha$ from an improved measurement of the fine structure. 

With the possibility of substantial improvement in the precision of
spectroscopic experiments (trapped atoms), a better determination of
$\alpha$ from measurement of the fine structure 
might be within reach in the near future.
Such a determination of the fine structure constant $\alpha$,
from the effect on which its name is based, would
complement other high precision determinations from solid state
physics and the anomalous magnetic moment of the electron.

We note that there are deviations of experimental 
data for excited nS--nP transitions from theory by more than one 
standard deviation but less than two standard deviations 
(see \cite{pipkin1} and references therein).
However, both theory of the Lamb shift
and spectroscopic techniques have improved since the
measurements were made, so one might expect a more precise comparison
of theory and experiment in the future.
The present uncertainty in the theory would in principle allow a 
determination of the fine structure constant with a relative 
uncertainty 
of less than 5 parts in $10^{9}$.  However at this level of precision
additional theoretical work might be needed to address questions
such as asymmetries in the natural line shape.
We only mention that for excited states, an
experimental determination of the fine structure could be simplified
by the slower decay (narrower line width) of the higher excited P 
states \cite{bookbethe}.

\section*{Acknowledgments}

The authors thank DFG for continued support (contract no. SO333/1-2).
We would like to thank K. Pachucki and S. Karshenboim
for stimulating and helpful discussions, and J. Urban for 
carefully reading the manuscript. (PJM)
acknowledges the Alexander von Humboldt Foundation for 
continued support. We also wish to 
acknowledge support from BMBF and from the Gesellschaft f\"{u}r 
Schwerionenforschung.

%
%

\begin{table}[htb]
\begin{center}
\begin{tabular}{c|c|c} 
\rule[-3mm]{0mm}{8mm}
contribution & $3P_{1/2}$ &  $3P_{3/2}$ \\ \hline
\rule[-3mm]{0mm}{8mm}
$F_{\rm nq}$ & 
$-1.433010(1) + 248/405\,\ln\left(\epsilon/(Z \alpha)^2\right) + $
& $-1.433010(1) + 248/405\,\ln\left(\epsilon/(Z \alpha)^2\right)$ \\  
\rule[-3mm]{0mm}{8mm}
$F_{\delta y}$ & 
$0.922653(1) - 20/81\,\ln\left(\epsilon/(Z \alpha)^2\right)$ 
 & $0.629717(1) - 20/81\,\ln\left(\epsilon/(Z \alpha)^2\right)$ \\ 
\rule[-3mm]{0mm}{8mm}
$F_{\delta H}$ & 
$0.356318(1) - 73/324\,\ln\left(\epsilon/(Z \alpha)^2\right)$ 
& $0.333053(1) - 55/324\,\ln\left(\epsilon/(Z \alpha)^2\right)$ \\ 
\rule[-3mm]{0mm}{8mm}
$F_{\delta E}$ &
$0.0406519(1) + 1/36 \, \ln\left(\epsilon/(Z \alpha)^2\right)$ 
 & $ 0.013551(1) + 1/108 \, \ln\left(\epsilon/(Z \alpha)^2\right)$ \\  
\rule[-3mm]{0mm}{8mm}
$F_{\delta \phi}$ & 
$-0.830340(1) + 40/81\,\ln\left(\epsilon/(Z \alpha)^2\right)$
 & $ -0.236869(1) + 13/108\,\ln\left(\epsilon/(Z \alpha)^2\right)$ \\ \hline
\rule[-4mm]{0mm}{10mm}
sum &
$-0.94378(1) + 268/405\,\ln\left(\epsilon/(Z \alpha)^2\right)$ 
 & $-0.69356(1) + 148/405\,\ln\left(\epsilon/(Z \alpha)^2\right)$ 
\end{tabular}
\end{center}
\caption{\label{table3P} Contributions of relative order 
$(Z \alpha)^2$ to the low--energy part $F_L$ for the 
$3P_{1/2}$ and $3P_{3/2}$ states. }
\end{table}

%
%

\begin{table}[htb]
\begin{center}
\begin{tabular}{c|c|c} 
\rule[-3mm]{0mm}{8mm}
contribution & $4P_{1/2}$ &  $4P_{3/2}$ \\ \hline
\rule[-3mm]{0mm}{8mm}
$F_{\rm nq}$ & 
$-1.512220(1) + 229/360 \, \ln\left(\epsilon/(Z \alpha)^2\right) $
& $-1.512220(1) + 229/360 \, \ln\left(\epsilon/(Z \alpha)^2\right)$ \\  
\rule[-3mm]{0mm}{8mm}
$F_{\delta y}$ & 
$0.966398(1) - 23/90 \, \ln\left(\epsilon/(Z \alpha)^2\right)$ 
 & $0.662154(1) - 23/90 \, \ln\left(\epsilon/(Z \alpha)^2\right)$ \\ 
\rule[-3mm]{0mm}{8mm}
$F_{\delta H}$ & 
$0.364541(1) - 2891/11520 \, \ln\left(\epsilon/(Z \alpha)^2\right)$ 
& $0.342940(1) - 439/2304 \, \ln\left(\epsilon/(Z \alpha)^2\right)$ \\ 
\rule[-3mm]{0mm}{8mm}
$F_{\delta E}$ &
$0.0335504(1) + 13/768 \, \ln\left(\epsilon/(Z \alpha)^2\right)$ 
 & $ 0.012904(1) + 5/768 \, \ln\left(\epsilon/(Z \alpha)^2\right)$ \\  
\rule[-3mm]{0mm}{8mm}
$F_{\delta \phi}$ & 
$-0.850066(1) + 787/1440 \, \ln\left(\epsilon/(Z \alpha)^2\right)$
 & $ -0.236345(1) + 53/288 \, \ln\left(\epsilon/(Z \alpha)^2\right)$ \\ \hline
\rule[-4mm]{0mm}{10mm}
sum &
$-0.99780(1) + 499/720 \, \ln\left(\epsilon/(Z \alpha)^2\right)$ 
 & $-0.73057(1) + 137/360 \, \ln\left(\epsilon/(Z \alpha)^2\right)$ 
\end{tabular}
\end{center}
\caption{\label{table4P} Contributions of relative order 
$(Z \alpha)^2$ to the low--energy part $F_L$ for the 
$4P_{1/2}$ and $4P_{3/2}$ states }
\end{table}

%
%

\begin{table}
\begin{center}
\begin{tabular}{c|r|r} 
\rule[-3mm]{0mm}{8mm}
contribution & $3P_{1/2}$ in kHz & $3P_{3/2}$ in kHz \\ \hline
\rule[-3mm]{0mm}{8mm}
one-loop self-energy & $-3477.349(5)$ & $4046.413(5)$ \\ 
\rule[-3mm]{0mm}{8mm}
two-loop self-energy & $7.705(23)$ & $-3.782(23)$ \\ 
\rule[-3mm]{0mm}{8mm}
three-loop self-energy & $-0.064$ & $0.032$ \\ 
\rule[-3mm]{0mm}{8mm}
vacuum polarization & $-0.122$ & $-0.027$ \\ 
\rule[-3mm]{0mm}{8mm}
$(Z \alpha)^4$ recoil & $0.641$ & $-0.320$ \\ 
\rule[-3mm]{0mm}{8mm}
$(Z \alpha)^5$ recoil & $-4.705(13)$ & $-1.915(13)$ \\ 
\rule[-3mm]{0mm}{8mm}
$(Z \alpha)^6$ recoil & $0.139$ & $0.139$ \\ \hline
\rule[-3mm]{0mm}{8mm}
Sum for $3P$ & $-3473.75(3)$ & $4037.75(3)$  
\end{tabular}
\end{center}
\caption{\label{lamb3p} Contributions to the Lamb shift in kHz for the 
$3P_{1/2}$ and $3P_{3/2}$ states.}
\end{table}

%
%

\begin{table}
\begin{center}
\begin{tabular}{c|r|r} 
\rule[-3mm]{0mm}{8mm}
contribution & $4P_{1/2}$ in kHz & $4P_{3/2}$ in kHz \\ \hline
\rule[-3mm]{0mm}{8mm}
one-loop self-energy & $-1403.102(2)$ & $1770.887(2)$ \\ 
\rule[-3mm]{0mm}{8mm}
two-loop self-energy & $3.252(10)$ & $-1.594(10)$ \\ 
\rule[-3mm]{0mm}{8mm}
three-loop self-energy & $-0.027$ & $0.014$ \\ 
\rule[-3mm]{0mm}{8mm}
vacuum polarization & $-0.054$ & $-0.012$ \\ 
\rule[-3mm]{0mm}{8mm}
$(Z \alpha)^4$ recoil & $0.270$ & $-0.135$ \\ 
\rule[-3mm]{0mm}{8mm}
$(Z \alpha)^5$ recoil & $-1.915(5)$ & $-1.915(5)$ \\ 
\rule[-3mm]{0mm}{8mm}
$(Z \alpha)^6$ recoil & $0.061$ & $0.061$ \\ \hline
\rule[-3mm]{0mm}{8mm}
Sum for $4P$ & $-1401.52(1)$ & $1767.30(1)$  
\end{tabular}
\end{center}
\caption{\label{lamb4p} Contributions to the Lamb shift in kHz for the 
$4P_{1/2}$ and $4P_{3/2}$ states. }
\end{table}

%
%
%

%
%
%

%
%
%

%
%

\end{document}